\documentclass{article}

\usepackage[margin=1in]{geometry}

\usepackage{amssymb}
\usepackage{amsmath}
\usepackage{booktabs}
\usepackage{verbatim}
\usepackage{enumitem}
\usepackage{subcaption}
\usepackage[colorlinks,citecolor=blue,urlcolor=blue,filecolor=blue,backref=page]{hyperref}
\usepackage{graphicx}

\usepackage{tikz}
\usetikzlibrary{bayesnet}

\newcommand\blfootnote[1]{%
\begingroup
\renewcommand\thefootnote{}\footnote{#1}%
\addtocounter{footnote}{-1}%
\endgroup
}

\RequirePackage{amsthm,amsmath,amsfonts,amssymb}
\RequirePackage[authoryear]{natbib}
\RequirePackage[colorlinks,citecolor=blue,urlcolor=blue]{hyperref}
\RequirePackage{graphicx}
\usepackage{multicol}

\usepackage[utf8]{inputenc}
\usepackage{float}
\usepackage{enumitem}
\usepackage{graphicx}
\usepackage{booktabs}
\usepackage[toc,page]{appendix}
\usepackage{caption}
\usepackage{bbm}
\usepackage{subcaption}
\usepackage{xcolor}

\graphicspath{{./figs/}}

\theoremstyle{remark}

\newcommand{\diag}{{\rm Diag}}

\usepackage{pifont}
\newcommand{\cmark}{\ding{51}}%
\newcommand{\xmark}{\ding{55}}%
\newcommand{\ml}[1]{{\fontfamily{pcr}\selectfont #1}}

\begin{document}

\title{Latent Community Adaptive Network Regression}

\author{ Heather Mathews$^\star$ and Alexander Volfovsky$^\dagger$ }
\date{Department of Statistical Science, Duke University}
\maketitle
\blfootnote{$^\star$heather.mathews@duke.edu, $^\dagger$alexander.volfovsky@duke.edu}
\begin{abstract}
The study of network data in the social and health sciences frequently concentrates on two distinct tasks (1) detecting community structures among nodes and (2) associating covariate information to edge formation. In much of this data, it is likely that the effects of covariates on edge formation differ between communities (e.g. age might play a different role in friendship formation in communities across a city). In this work, we introduce a latent space network model where coefficients associated with certain covariates can depend on latent community membership of the nodes. We show that ignoring such structure can lead to either over- or under-estimation of covariate importance to edge formation and propose a Markov Chain Monte Carlo approach for simultaneously learning the latent community structure and the community specific coefficients. We leverage efficient spectral methods to improve the computational tractability of our approach.
\end{abstract}
\section{Introduction}

Network data provide a unique opportunity to study the patterns of societal interaction. The inferential task in the social and health sciences often involves learning about the influence that individual and group level information can have on the formation of network ties, such as friendships among students within and across schools (e.g. the  PROSPER \citep{PROSPER} and National Longitudinal Study of Adolescent to Adult Health \citep{addHealth2009}). A common explanation for the formation of connections among individuals is the notion of homophily---that birds of a feather flock together \citep{ShrumFriend1988,genderDiff2005}.
This idea has driven the development of two broad classes of statistical models: (i) community detection models that attempt to identify important and similar groups of individuals in the graph \citep{Holland83} and (ii) more general network regression models that postulate a direct link between observed individual attributes or covariates and the observed network interactions \citep{hoff2002latent,hoff2018amen,Holland81}.  Several approaches have brought the two tasks closer together by conjecturing that under some distributional assumptions, covariate information can assist in detecting communities in the network \citep{Rohe17,Mele19}.

An important observation that motivates the development in this paper is that the differences between communities within networks likely translate to differences in the impact of certain covariates for explaining edge formation. The degree corrected stochastic blockmodel was one of the first models that allows within community heterogeneity such that communities are not just detected based on degree \citep{qin2013regularized}. Importantly, such differences are frequently observed empirically. 
Examples of such behavior abound in the literature: \citet{aukett1988gender} shows that prediction of same sex friendships differ between men and women: women typically base relationships on sharing emotions and discussing personal issues whereas men tend to build relationships based on activities that they do together. \citet{Staber2013} studies how men and women form entrepreneurial relationships differently, finding that women tend to form larger networks with more male ties and more strangers. \citet{Chamberlain2007} describes classmate relationship formation between children with and without autism in a mixed environment. \citet{bail2018exposure} shows differential effects of opposite party exposure on polarization. For all of these examples, the community labels are known, and, given this information, differences between the communities and effects of covariates are easy to evaluate.

Unlike the aforementioned studies, community labels are frequently unavailable but differences in covariate influence are still likely to persist via latent or unobserved communities.
The literature lacks methods that address the important question of varying influence of covariate information based on latent community membership of individuals in a network. In this paper, we propose a network model that addresses this gap by leveraging learned latent community information to better describe the effects of certain covariates on friendship formation. 

We motivate our development by studying the high school friendship networks of the National Longitudinal Study of Adolescent Health (AddHealth), a nationally representative study conducted during the 1994-1995 school year. In AddHealth,  high school students nominate their top 5 male and top 5 female friends. If student $i$ nominates student $j$, then a directed edge exists from $i$ to $j$. Covariate information on students was collected including grade, smoking status, drinking habits, club involvement, GPA, and sport involvement \citep{addHealth2009}. Many of these covariates have been shown to be predictors of friendship formation \citep{drinkPressure2010,hoff2013amen}. Given the high school setting, it is however probable that latent communities that are not associated with observed covariates are influencing friendships as well. While the communities are not predicted by observed covariates, the ties within them likely are influenced by covariates in different ways.

\subsection{The AddHealth data}\label{intro_ah}
\begin{figure}[H]
	\centering
	\begin{minipage}[t]{\dimexpr.5\columnwidth-1em}
		\centering
       \includegraphics[width=\columnwidth]{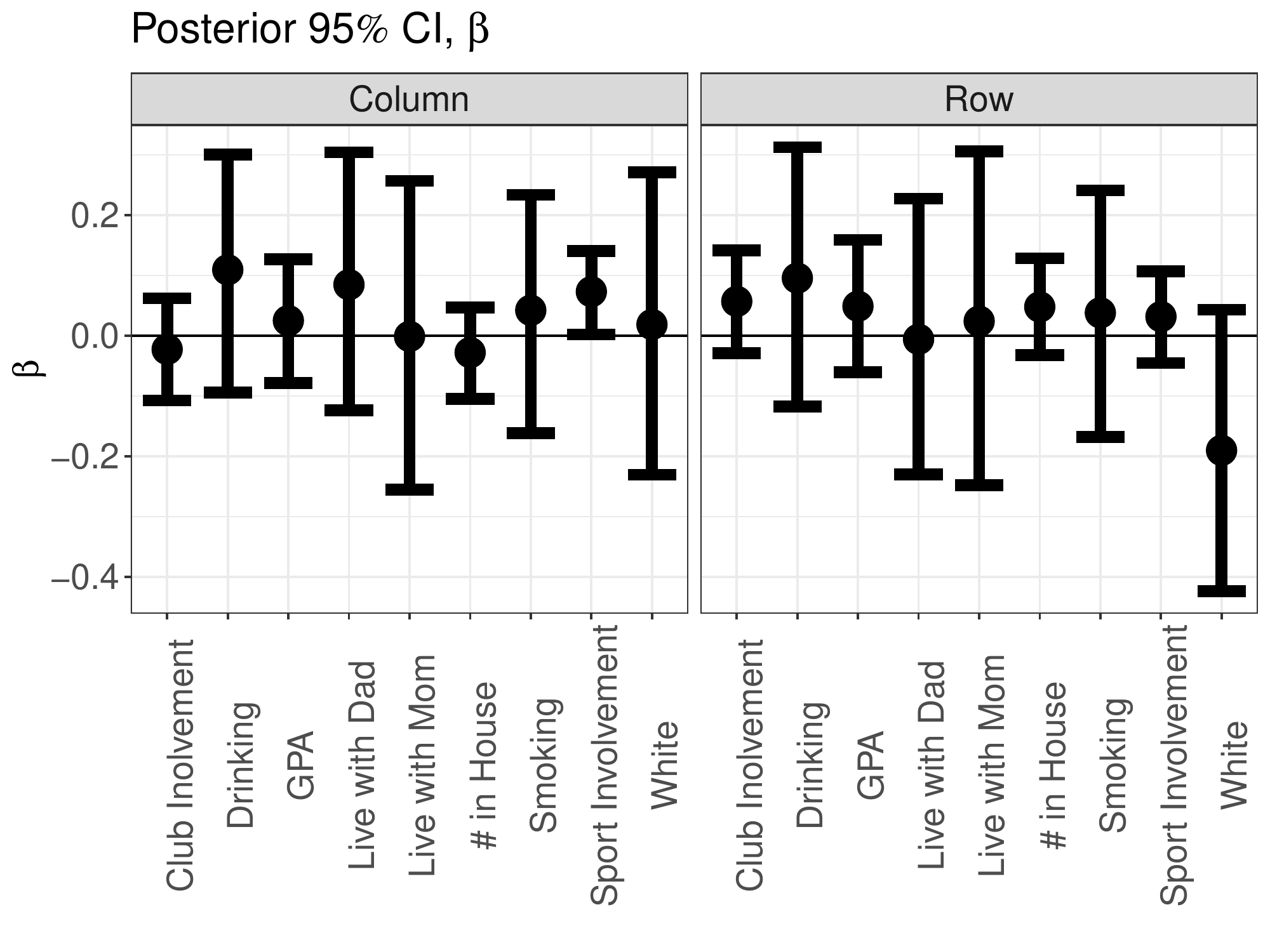}
    	\caption{$95\%$ CI for $\beta$ estimates when fitting AMEN model on AddHealth network \label{fig:amenR2_22}}
	\end{minipage}\hfill
	\begin{minipage}[t]{\dimexpr.5\columnwidth-1em}
		\centering
		\includegraphics[width=\columnwidth]{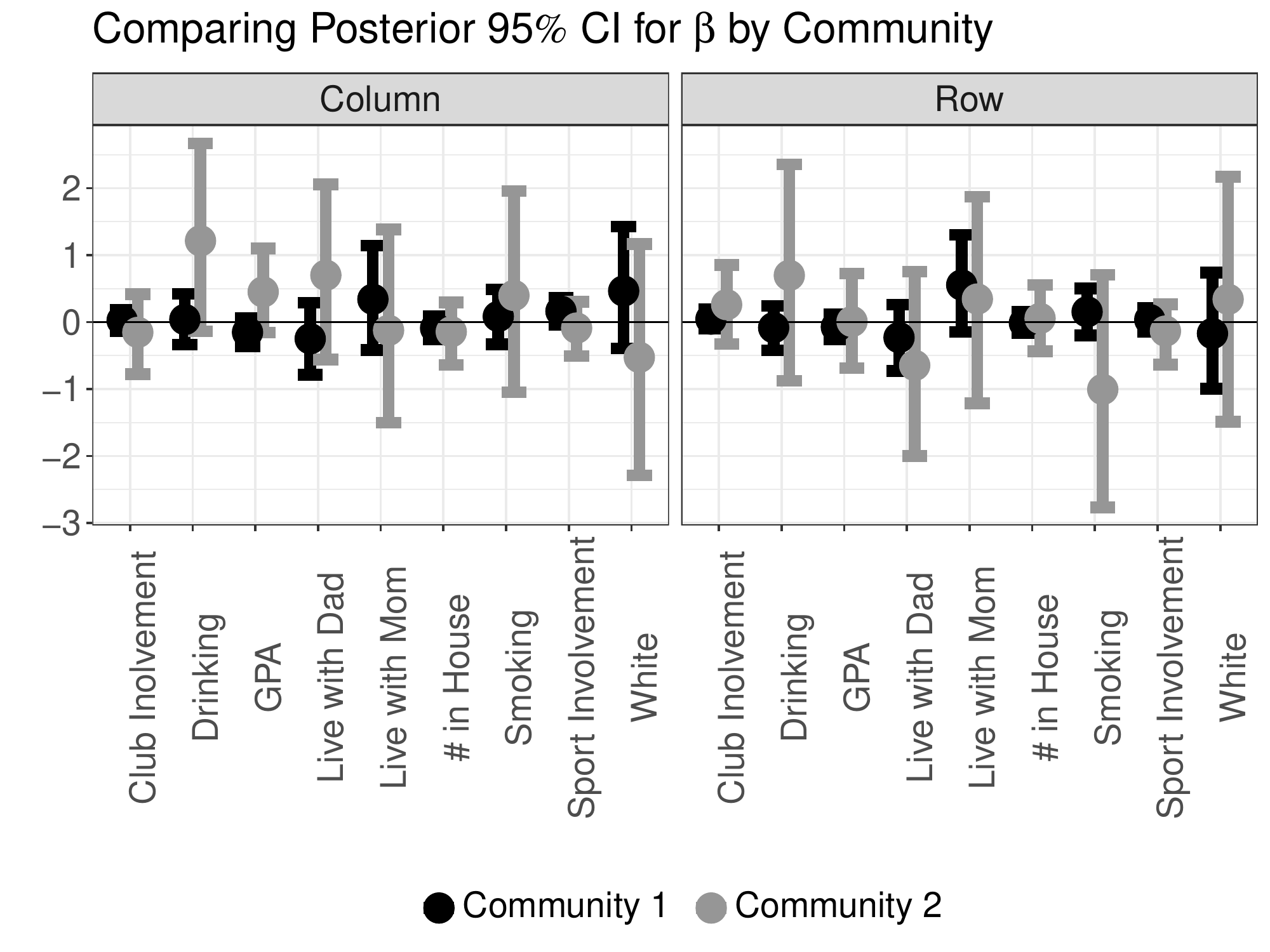}
		\caption{Running separate AMEN models on school network partitioned by initial estimated latent communities }
		\label{fig:sepAmen}
	\end{minipage}
\end{figure}

We motivate the development of this paper by studying friendship formation in one AddHealth school. We first analyze the data using a flexible but not community driven network model called the Additive and Multiplicative Effects Network (AMEN) model. In this model covariates are linearly associated with the probability of edge formation, with positive coefficients meaning that there is a higher probability of a friendship. Covariate information for both the sender and the receiver of the ties can influence the formation of a directed edge. Figure~\ref{fig:amenR2_22} presents 95\% posterior credible intervals for sender (labeled row) and receiver (labeled column) effects.

The results indicate that being involved in sports potentially increases individuals' sociability (row effect) and popularity (column effect). While the model appears to fit fairly well according to posterior predictive checks, standard community detection tools detect communities that are not directly correlated with any predictor. As such, we cannot rule out that the covariates have different effects on friendship formation within these communities. We explore this possibility by naively fitting simple AMEN models within each of the found communities. Figure \ref{fig:sepAmen} depicts the results of a naive application of community detection on the original network and then performing separate fits of AMEN within each community. We see that individuals in different estimated communities might be influenced differently by certain covariates such as being white and club involvement. However, this two-step procedure has clear downsides: partitioning the data leads to a large loss of information since only connections between people within the same estimated community are studied and the uncertainty in the community membership estimation is not propagated throughout the analysis.  In the next section we propose a joint model that resolves these issues.
\subsection{Notation and model setup}
To fix notation, we represent a directed network with a $n \times n$ matrix $Y$ (frequently referred to as a sociomatrix) that describes the relationships between $n$ nodes. The entry of $y_{ij}$ is a binary indicator as to whether or not an edge exists between person $i$ and person $j$. We further observe $p$ covariates associated with these nodes as senders (row covariates) and as receivers (column covariates). For notational simplicity, assume that the number of row and column covariates is the same (however this is not required). Below we generalize the AMEN model that incorporates covariate coefficient estimation along with latent multiplicative and nodal random effects \citep{hoff2013amen}. Specifically, we consider the generative model where $y_{ij} = \mathbbm{1}_{z_{ij} > 0}$ (that is, the observed edge $y_{ij}$ is a function of a latent unobserved strength of connection $z_{ij}$) where:
\begin{equation}\label{eq:OGamen}
 z_{ij} = \beta_0  + \sum_{l=1}^p  (x_{rli}\beta_{rl} + x_{clj}\beta_{cl}) + \mathbf{u}_i^t\Lambda \mathbf{v}_j + a_i+b_j + \epsilon_{ij}.
 \end{equation}
 In matrix form, this relational latent representation of $Y$ can be written as:
 \begin{equation}\label{eq:amen_basic}
 Z = \beta_0 \mathbf{11}^t + \sum_{l=1}^p  (X_{rl}\beta_{rl} + X_{cl}\beta_{cl}) + U\Lambda V^t + \mathbf{a} \mathbf{1}^t+\mathbf{1} \mathbf{b}^t + E
 \end{equation}
 where $E$ is an $n \times n$ matrix of possibly correlated errors. 
The flexibility of this model is in the specification of the random effects. 
Row ($a$) and column ($b$) additive effects capture sociability and popularity of nodes, respectively (this generalizes the degree corrected stochastic blockmodel) \citep{qin2013regularized}. 
Since a node's popularity and sociability are likely correlated, we specify them jointly,
\begin{center}
$\begin{pmatrix}
a_i\\b_i \end{pmatrix} \stackrel{i.i.d}{\sim}  N(0, \Sigma_{ab})
$
with $\Sigma_{ab} = \begin{bmatrix}
\sigma^2_a & \sigma_{ab}\\
\sigma_{ab} & \sigma^2_b
\end{bmatrix}$
\end{center}
where $\sigma^2_a$, $\sigma^2_b$ represent the variance among row and column means and $\sigma_{ab}$ represents the covariance between row and column effects.
In the standard AMEN model,  $U\Lambda V^t$ captures potential higher order latent dependencies between nodes. For example, if there are homophilous latent communities in $Y$ and node $i$ and $j$ belong to the same community, $\mathbf{u}_{i}^t
\Lambda \mathbf{v}_{j}$ would increase $z_{ij}$. In the analysis of Section \ref{intro_ah}, latent multiplicative dimension is equal to 2, that is, $\mathbf{u}_i,\mathbf{v}_j\in\mathbb{R}^2$. We note that the model is overparametrized when $\Lambda$ is included without a restriction on the elements of $\mathbf{u_i}$ and $\mathbf{v_j}$. As we are about to introduce such a restriction we maintain such a notation. Furthermore, this formulation showcases how the AMEN framework generalizes both the SBM and the latent distance model \citep{hoff2008modeling}. This also allows the AMEN model to capture information about triads without requiring the explicit inclusion of those statistics as in an ERGM. 
To capture the likely reciprocity between directed friendships, one can introduce dyadic dependence in $Z$ by letting $\epsilon_{ij}$ and $\epsilon_{ji}$ be correlated: 
 \begin{center}
 $\begin{pmatrix}
\epsilon_{ij}\\
\epsilon_{ji}
\end{pmatrix} \stackrel{i.i.d}{\sim} N(0, \Sigma_{\epsilon})$ where $\Sigma_{\epsilon} = \begin{bmatrix}
1 & \rho \\ 
\rho & 1
\end{bmatrix}$
\end{center} where $\rho$ captures this dyadic correlation \citep{hoff2018amen}. 
The parameters of the standard AMEN model in Eq~\eqref{eq:OGamen} are estimated using a Markov Chain Monte Carlo (MCMC) algorithm. 
In this paper, we extend the AMEN model by considering the connection between the coefficients $\beta$ and the multiplicative effects $U, V$.

\section{Extending the AMEN Model}\label{fullModel}
When latent communities exist, connections in different communities may be influenced by covariates in different ways. We present a model in which latent community detection is combined with network regression by allowing for community dependent coefficients. We concentrate on discrete communities, considering the previously defined multiplicative variables to be $K$-dimensional binary vectors where $u_{ik} = 1$ if unit $i$ is in community $k$ (assume $\sum_{k = 1}^K u_{ik}= 1$) as a sender and similarly $v_{ik}$ for a receiver, where $\Lambda \in \mathbb{R}^{K \times K}$ represents the relationships between communities. One of the important upsides of this formulation is the interpretability of the parameters due to the discrete community memberships of each individual. We discuss a natural extension to the mixed membership and general latent space model that loses this interpretability in Section \ref{discussion}.
\subsection{The Model}\label{ubetatilde}
In order to allow for community dependent coefficients we re-index them as a function of $U$ and $V$. That is, let
\begin{equation}
	z_{ij} = \beta_0 + \sum_{l=1}^p  (x_{rli}\beta_{rl f(\mathbf{u}_i)}+ x_{clj}\beta_{cl f(\mathbf{v}_j)}) + \mathbf{u}_i^t\Lambda \mathbf{v}_j + a_i+b_j + \epsilon_{ij}.
	\label{eq:amen_dep}
\end{equation}
where $f(\mathbf{u}_i) = \sum_{k =1 }^K k \times u_{ik}$. To facilitate computation we derive a matrix formulation for Eq~\eqref{eq:amen_dep}.
To allow each latent community to have its own covariate coefficient, let $\tilde{\beta}_{r}$, $\tilde{\beta}_{c}$ denote $p \times K$ matrices of coefficients for row and column covariates respectively. Altering Eq~\eqref{eq:OGamen} to accommodate the community dependent covariate coefficients yields the following:
\begin{equation}
Z = \beta_0 \mathbf{11}^t + \sum_{l=1}^p (\diag(\tilde{\boldsymbol\beta}_{rl}U^t) X_{rl} + X_{cl} \diag(\tilde{\boldsymbol\beta}_{cl}V^t)) + U\Lambda V^t + \mathbf{a 1}^t+\mathbf{1 b}^t + E.
\label{eq:amen_dep_matrix}
\end{equation}
To see that this corresponds to Eq~\eqref{eq:amen_dep}, consider the following three community example. Nodes 1-4 belong to communities 1, 2, 1, and 3 respectively.
\begin{center}
	$\tilde{\boldsymbol\beta}_{rl} U^t = \begin{bmatrix}
	\textcolor{gray}{\tilde{\beta}_{rl,1}} & \textcolor{darkgray}{\tilde{\beta}_{rl,2}} & \tilde{\beta}_{rl,3} 
	\end{bmatrix}
	\begin{bmatrix}
	1 & 0 & 1 &0 \\
	0 & 1  & 0 & 0 \\ 
	0&  0  & 0 & 1
	\end{bmatrix}$
\end{center}
\begin{center}
	$  \diag(\tilde{\boldsymbol\beta}_{rl}U^t)  X_{r} =  
	\begin{bmatrix} 
		\textcolor{gray}{\tilde{\beta}_{rl,1}} & 0 & 0& 0\\
	0& \textcolor{darkgray}{\tilde{\beta}_{rl,2}}&0 & 0\\
	0&0&	\textcolor{gray}{\tilde{\beta}_{rl,1}}& 0\\
	0 &0&0& \tilde{\beta}_{rl,3}
	\end{bmatrix}
	\begin{bmatrix}
	\textcolor{gray}{x_1}  & \textcolor{gray}{x_1}  & \textcolor{gray}{x_1}  &\textcolor{gray}{x_1} \\
	\textcolor{darkgray}{x_2} & \textcolor{darkgray}{x_2} & \textcolor{darkgray}{x_2}  & \textcolor{darkgray}{x_2}\\
	\textcolor{gray}{x_3} &\textcolor{gray}{x_3} &  \textcolor{gray}{x_3}  &\textcolor{gray}{x_3}\\
	x_4 & x_4 & x_4& x_4
	\end{bmatrix} = 
		\begin{bmatrix}
	\textcolor{gray}{\tilde{\beta}_{rl,1}}\textcolor{gray}{x_1}  & \textcolor{gray}{\tilde{\beta}_{rl,1}}\textcolor{gray}{x_1}  &\textcolor{gray}{\tilde{\beta}_{rl,1}} \textcolor{gray}{x_1}  &\textcolor{gray}{\tilde{\beta}_{rl,1}}\textcolor{gray}{x_1} \\
\textcolor{darkgray}{\tilde{\beta}_{rl,2}}\textcolor{darkgray}{x_2} & \textcolor{darkgray}{\tilde{\beta}_{rl,2}}\textcolor{darkgray}{x_2} & \textcolor{darkgray}{\tilde{\beta}_{rl,2}}\textcolor{darkgray}{x_2}  & \textcolor{darkgray}{\tilde{\beta}_{rl,2}}\textcolor{darkgray}{x_2}\\
\textcolor{gray}{\tilde{\beta}_{rl,1}}\textcolor{gray}{x_3} &\textcolor{gray}{\tilde{\beta}_{rl,1}}\textcolor{gray}{x_3} &  \textcolor{gray}{\tilde{\beta}_{rl,1}}\textcolor{gray}{x_3}  &\textcolor{gray}{\tilde{\beta}_{rl,1}}\textcolor{gray}{x_3}\\
\tilde{\beta}_{rl,3}	x_4 & \tilde{\beta}_{rl,3}x_4 &\tilde{\beta}_{rl,3} x_4& \tilde{\beta}_{rl,3}x_4

	\end{bmatrix}
	$.
\end{center}
To construct the above for column covariates, replace  $  \diag(\tilde{\boldsymbol\beta}_{rl}U^t)  X_{r}$ with $X_{c} \diag(\tilde{\boldsymbol\beta_{cl}}V^t)$. Dyadic covariates can be represented similarly, and this discussion is relegated to Supplement S2. 

\subsection{Relationship between $\beta$ in Eq~\eqref{eq:com_dep} and $\tilde{\beta}$ in Eq~\eqref{eq:com_indep}}\label{asymptoticB}
In this section we describe the way in which our model generalizes AMEN and what to expect from fitting the simpler model when the truth is community dependent. For the purpose of this example only, consider a setting where continuous interactions between units are observed, community memberships of units are known, there is no reciprocal correlation, and we are only concerned with a single potential row predictor. In this case, there are two possible model formulations: 
\begin{eqnarray}
\textrm{community dependent model:\quad} & Z =  \diag(\tilde{\boldsymbol\beta}U^t) X+ E,\label{eq:com_dep}\\
\textrm{independent model:\quad} & Z = X\beta + E,\label{eq:com_indep}
\end{eqnarray}
where $\epsilon_{ij} \sim N(0, 1)$ and $p = 1$. Note that this can easily be extended to a correlated error structure by using the quantities described in Section \ref{posteriors} as well as be extended to include multiple row and column covariates. Let $Z^{(vec)}$ be the vectorized version of $Z$ and $X^{(vec)}$ the vectorized version of $X$ \citep{kolda2009tensor}. We can write the OLS estimates of $\beta$ and $\tilde{\boldsymbol\beta}$ as \begin{center}
	$\hat{\beta} = \frac{\sum_{i = 1}^{n^2} x_i z_i }{\sum_{i = 1}^{n^2} x_i^2 }$ ,\quad 
	$\hat{\tilde{\beta}}_{k} = \frac{\sum_{i \in c_k} H_{ik} z_i }{\sum_{i \in c_k} H_{ik}^2},$
\end{center}
where $c_k$ is the set of all indices for individuals in community $k$, $H = \{\mathbf{1}_n \otimes [I_n \circ X] U\}$, where ``$\otimes$'' denotes the Kronecker product and ``$\circ$" denotes the Hadamard product. 
The numerator of the expression for $\hat\beta$ can be rewritten as $\sum_{i = 1}^{n^2} x_i z_i = \sum_{k = 1}^K \sum_{i \in c_k} x_i z_i = \sum_{k = 1}^K \sum_{i \in c_k} H_{ik} z_i$. Combined with the identity $X^{(vec)} = \sum_{k=1}^K H_{\cdot k}$, we can express $\hat{\beta}$ in terms of $\hat{\tilde{\boldsymbol\beta}}$:

\begin{equation}
\hat{\beta} = \frac{(\sum_{k = 1}^K \hat{\tilde{\beta}}_{k} (\sum_{i \in c_k} H_{ik}^2 ))}{\sum_{i = 1}^{n^2} x_i^2 } = \frac{ n\sum_{k=1}^K S_k^2 \hat{\tilde{\beta}}_k |c_k|}{n^2 S^2}
\end{equation}
where $S^2_k =\frac{1}{|c_k| n} \sum_{i \in c_k} H_{ik}^2$ is the sample variability of $X$ for each observed community, and $S^2 =  \frac{1}{n^2} \sum_{i=1}^{n^2} x_i ^2$ is the overall variability. 
The connection between the two models is clear --- the community independent model represents a weighted average of the community dependent coefficients. As such, we can translate between the community dependent and independent models, making over-specification less problematic. We demonstrate this behavior empirically for the full model in Section~\ref{simulations}.

.
\section{Bayesian Computation}\label{bayes}
We rarely observe continuous interactions between individuals and rather have binary or ranked values. This makes maximum likelihood inference much more challenging, but tractable Bayesian inference is feasible. In this section, we derive a Metropolis within Gibbs sampler for the parameters of the model proposed in Eq~\eqref{eq:amen_dep_matrix}. Our derivations generalize those presented in \citet{hoff2018amen} by incorporating community dependence. In the remainder of this paper we consider the sender and receiver communities to be the same (that is $U = V$). This is a reasonable assumption for much of social science data. Generalizing to $U \neq V$ is a simple exercise that follows the same steps outlined below.  The Markov Chain Monte Carlo sampler for sampling from the posterior $p(\tilde\beta, Z, \mathbf{a}, \mathbf{b}, U, \Lambda, \rho | Y, X)$ is given by:

\begin{enumerate}
    \item Update $\Lambda$: Sample $\Lambda^{(s + 1)} \sim N(V_{\Lambda}m_{\Lambda}, V_{\Lambda})$ where $V_{\Lambda}$, and $m_{\Lambda}$ depend on $U^{(s)}$,  $\tilde{\beta}^{(s)}$, $Z^{(s)}$, $\mathbf{a}^{(s)}$, $\mathbf{b}^{(s)}$, and  $\rho^{(s)}$.
    \item Update $U$ and $Z$:
    \begin{enumerate}
        \item With uniform probability, randomly select a unit $i$ to update. Propose new membership $\mathbf{u}^\star_{i\cdot}$.
        \item Compute the acceptance probability 
        $$r = \min \Bigg(1,\frac{P(\mathbf{u}^\star_i | Y, U_{-i}^{(s)}, \mathbf{a}^{(s)}, \mathbf{b}^{(s)}, \tilde{\beta}^{(s)}, \Lambda^{(s+1)}, \rho^{(s)})}{P(\mathbf{u}_i | Y,U_{-i}^{(s)}, \mathbf{a}^{(s)}, \mathbf{b}^{(s)}, \tilde{\beta}^{(s)}, \Lambda^{(s+1)}, \rho^{(s)})}\Bigg), $$
        which does not depend on any $Z$.
        \item Sample $w \sim uniform(0,1)$ and if $w < r$ then we accept and set $\mathbf{u}_{i\cdot}^{(s + 1)} = \mathbf{u}^\star_{i\cdot}$, otherwise $\mathbf{u}_{i\cdot}^{(s + 1)} = \mathbf{u}_{i\cdot}^{(s)}$. If the proposal is accepted, sample a new $Z$ directly: $\{\mathbf{z}^\star_{i,\cdot},\mathbf{z}^\star_{\cdot,i}\} \sim Normal(\mu_z, \Sigma_{z})$ where the parameters depend on $U^{(s+1)}$, $\Lambda^{(s + 1)}$, $\rho^{(s)}$ $\tilde{\beta}^{(s)}$, $Z_{-i\cdot}^{(s)}$,$\mathbf{a}^{(s)}$,$\mathbf{b}^{(s)}$.
    \end{enumerate}
    \item Update $\rho$: Sample $\rho^{(s+1)}$ using a Metropolis Hastings update. This is based off of the full conditional that depends on $U^{(s+1)}$, $\Lambda^{(s + 1)}$, $\rho^{(s)}$, $\tilde{\beta}^{(s)}$,  and $Z^{(s)}$.
    \item Update $\tilde{\beta}$: Sample $\tilde{\beta}^{(s + 1)}$ from $N(Vm, V)$ where $V, m$ depend on $U^{(s +1)}$,
    $Z^{(s)}$, $\mathbf{a}^{(s)}$,$\mathbf{b}^{(s)}, \rho^{(s + 1)}$
    \item Update $Z$ (again): Sample $T = \textrm{lower tri}(Z)|\textrm{upper tri}(Z^{(s)})$ and set $\textrm{lower tri}(Z^{(s+1)}) = T$ and then sample $T = \textrm{upper tri}(Z)|\textrm{lower tri}(Z^{(s+1)})$ and set $\textrm{upper tri}(Z^{(s+1)}) = T$.
     \item Update $\mathbf{a},\mathbf{b}$: Sample $\{\mathbf{a},\mathbf{b}\}^{(s + 1)}$ from a multivariate normal where the parameters depend on $U^{(s+1)}$,  $ \Lambda^{(s + 1)} $, $\rho^{(s + 1)} $, $\tilde{\beta}^{(s + 1)}$, $\Sigma_{ab}^{(s)}$, and $Z^{(s + 1)}$.
    \item Update $\Sigma_{ab}$: Sample $\Sigma_{ab}^{(s + 1)}$ from a conjugate Inverse-Wishart distribution that depends on $Z^{(s + 1)}$, $\mathbf{a}^{(s + 1)}$ and $\mathbf{b}^{(s + 1)}$.
\end{enumerate}
\subsection{Priors}\label{priors}
Normal priors are placed on $\beta$ related coefficients:
\begin{center}
 $ \beta_0 \sim N(\mu_{\beta_0}, \sigma^2_{\beta_0})$, $  \tilde{\boldsymbol\beta}_{rl},\tilde{\boldsymbol\beta}_{cl} \sim N(\boldsymbol\mu_{\tilde{\beta}}, \Sigma_{\tilde{\beta}})$.\end{center}
 Note that $\boldsymbol\mu_{\tilde{\beta}}$ is a $K$ dimensional vector and $\Sigma_{\tilde{ \beta}}$ is a $K \times K$ matrix.  For $\Lambda$, a normal prior is also used,  $ vec(\Lambda) \sim N(\boldsymbol\mu_{\Lambda}, \Sigma_{\Lambda})$ where $\boldsymbol\mu_{\Lambda}$ is a $K^2$ vector (if made symmetric, $K(K-1)/2$). For $U$, assume that each person has equal probability of belonging to community $k \in \{1,...,K\}$. As such, place a uniform distribution over the canonical basis for $U$. The remaining priors for variance and covariance parameters are default choices (Gamma and Inverse-Wishart) as in \citet{hoff2018amen}. An arc sine prior is used for $\rho$ and an Inverse-Wishart$(I_{2}, 4)$ is used for $\Sigma_{ab}$.
\subsection{Computation}\label{posteriors}
 This section provides the calculations needed to sample the posterior of the parameters of interest. 
 \subsubsection{Updating $\tilde{\beta}$, $\mathbf{a}$, $\mathbf{b}$}
 For ease of notation, let $p = 1$ for the derivations below. 
 In order to derive full conditionals for $\tilde{\boldsymbol\beta}_r, \tilde{\boldsymbol\beta}_c$, define 
 \begin{align*}
 \tilde{Z_r} &= Z-\beta_0 \mathbf{11}^t -  X_{c}\diag(\tilde{\boldsymbol\beta_c}V^t) - U\Lambda V^t - \mathbf{\mathbf{a1}}^t- \mathbf{\mathbf{1b}}^t  = Diag(\tilde{\boldsymbol\beta_r} U^t)X_r +E\\
\tilde{Z_c} &= Z- \beta_0 \mathbf{11}^t - \diag(\tilde{\boldsymbol\beta_r}U^t)X_r - U\Lambda V^t - \mathbf{\mathbf{a1}}^t- \mathbf{\mathbf{1b}}^t
 = X_cDiag(\tilde{\boldsymbol\beta_c}V^t) +E.
 \end{align*}
Throughout, the computational complication in the full conditionals comes from the correlation between $z_{ij}$ and $z_{ji}$. To address this, we can create decorrelated versions of these as 
 \begin{equation}
 \tilde{Z}_{c*} = s\times \tilde{Z}_c + t\times \tilde{Z}_c^t \quad \textrm{and} \quad
 \tilde{Z}_{r*} = s\times \tilde{Z}_r + t\times \tilde{Z}_r^t.\label{eq:decor}
 \end{equation}
 where $s = \frac{\{(1 + \rho)^{-1/2} + (1-\rho)^{-1/2}\}}{2}$ and $t =\frac{\{(1 + \rho)^{-1/2} - (1-\rho)^{-1/2}\}}{2}.$ 
 It is possible to then vectorize each of the objects in Eq~\eqref{eq:decor} to perform a standard conjugate update, but this will require inverting $n^2\times n^2$ matrices. Instead, we are able to leverage the Kronecker structure of the predictors to describe the full conditional of $\tilde\beta_c$ and $\tilde\beta_r$ in terms of the following, easy to compute objects:
 \begin{equation}\label{eq:Hc}
 H_c = s\times \{[I_n \circ X_c]V \otimes \mathbf{1}_n\} + t \times \{\mathbf{1}_n \otimes [I_n \circ X_c]V\},
 \end{equation}
 \begin{equation}\label{eq:Hr}
 H_r = s\times\{\mathbf{1}_n \otimes [I_n \circ X_r]U\}+ t \times \{[I_n \circ X_r]U \otimes \mathbf{1}_n\}.
 \end{equation}
 The full conditionals for $\tilde{\boldsymbol\beta}_r$ and $\tilde{\boldsymbol\beta}_c$ are given by:
  \begin{align*}\label{eq:condBetaR}
 \tilde{\boldsymbol\beta}_r| X_{c}, X_{r}, U,V, \Lambda, Z, \tilde{\boldsymbol\beta}_c, \mathbf{a},\mathbf{b}\sim& N(V_rm_r, V_r) \textrm{ and } \\
\tilde{\boldsymbol\beta}_c| X_{c}, X_{r}, U,V, \Lambda, Z, \tilde{\boldsymbol\beta}_r, \mathbf{a},\mathbf{b}\sim& N(V_cm_c, V_c).
 \end{align*}
 where
 \begin{equation}
 \begin{split}
 V_c &= (H_c^t H_c + \Sigma_{\tilde{\boldsymbol\beta_c}}^{-1})^{-1}\\
 m_c &=vec(\tilde{Z}_{c*})^tH_c + \boldsymbol\mu_{\tilde{\boldsymbol\beta}_{c}} \Sigma_{\tilde{\beta}_c}^{-1}
 \end{split}
 \quad \quad
 \begin{split}
 V_r &= (H_r^t H_r + \Sigma_{\tilde{\boldsymbol\beta_r}}^{-1})^{-1}\\
 m_r &=vec(\tilde{Z}_{r*})^tH_r +\boldsymbol\mu_{\tilde{\beta}_{r}} \Sigma_{\tilde{\beta}_r}^{-1}
 \end{split}
 \end{equation}
 Updating $(\mathbf{a},\mathbf{b})$ follows similarly by computing $\tilde Z_{a}$ and $\tilde Z_{b}$, applying the transformation in Eq~\eqref{eq:decor} and proceeding as above.
 \subsubsection{Update $\Lambda$}
 
To update $\Lambda$ we leverage the conjugacy of the normal prior again. Write the de-meaned and decorrelated data as
 \begin{center}
      $\tilde{Z}_{\Lambda} = Z-\beta_0 \mathbf{11}^t - \diag(\tilde{\boldsymbol\beta_r}U^t)X_{r} -X_{c}\diag(\tilde{\boldsymbol\beta_c}V^t)- \mathbf{a1}^t- \mathbf{1b}^t $ and
 $\tilde{Z}_{\Lambda *} = s\times \tilde{Z}_{\Lambda}  +t\times \tilde{Z}_{\Lambda}^t $.
 \end{center}
 We can define the counterpart to Eq \eqref{eq:Hc}: 
 \begin{center}
   $H_{\Lambda} =   s \times (V \otimes U) + t \times (U \otimes V)$.
 \end{center}
The full conditional of $\Lambda$ is given by
  \begin{equation}
 \Lambda|X_c, X_r, U,V,  \tilde{\boldsymbol\beta}_c,\tilde{\boldsymbol\beta}_r,Z,\mathbf{a},\mathbf{b} \sim  N(V_{\Lambda}\mathbf{m}_{\Lambda}, V_{\Lambda})  \label{eq:condLambda}
\end{equation}
 where 
 $\mathbf{m}_{\Lambda} = vec(\tilde{Z}_{\Lambda *})^t H_{\Lambda}  + \mu_{\Lambda} \Sigma_{\Lambda}^{-1}$
 and 
 $V_{\Lambda} = (H_{\Lambda} ^t H_{\Lambda}  +\Sigma_{\Lambda}^{-1} )^{-1}.$
 
\subsubsection{Updating $U$ and $Z$}
Note that the update for different sending and receiving communities can be derived analogously. Updating large numbers of discrete parameters is notoriously associated with poor mixing of MCMC. Given the size of the parameter space, we propose to update only a small portion of $U$ at a time. We observe that this allows for better overall mixing. In the original AMEN model, when $U$ is a continuous object in $\mathbb{R}^d$ the update can naturally be implemented as a Gibbs step conditional on $Z$ \citep{hoff2018amen}. While updating $U|Z,\cdots$ is still possible using a Metropolis step when $U$ takes on discrete values, the autocorrelation in the $Z$'s induces an unacceptably high autocorrelation in the $U$s, thus again inhibiting mixing. We propose to sample $U|Y,\cdots$ marginializing out the $Z$s. This requires an evaluation of the bivariate probit likelihood. Details on the explicit form of this marginal bivariate probit likelihood are provided in Supplement S4. In order to facilitate the remaining updates (such as those in the previous section) we must re-sample the $Z$s --- this must now be done marginally as well since the marginalization of the $Z$s in the $U$ update changes the state space, meaning we cannot condition on previous $Z$s.

We introduce the following update: For a randomly selected unit $i$ we propose $\mathbf{u}^\star_{i\cdot}$ uniformly from $\{1, ...,K\}$ . If $\mathbf{u}^\star_{i\cdot}$ is accepted then $ \mathbf{z}_{i.}$ and  $ \mathbf{z}_{.i}$ are sampled from the bivariate truncated normal distribution unconditional on previous $Z$s. If $\mathbf{u}^\star_{i\cdot}$ is not accepted, then the $Z$s can be sampled using a Gibbs step. The acceptance probability for the $\mathbf{u}^\star_{i\cdot}$ is

$$\min \Bigg(1,\frac{P(\mathbf{u}^\star_i | Y, U_{-i},\mathbf{a}, \mathbf{b}, \tilde{\beta}, \Lambda, \rho)}{P(\mathbf{u}_i| Y, U_{-i},\mathbf{a}, \mathbf{b}, \tilde{\beta}, \Lambda,\rho)}\Bigg) $$  
We note that the acceptance ratio has had $Z$ marginalized out. We can then write
\begin{equation}
P(\mathbf{u}_{i}^*|Y, U_{-i},\mathbf{a}, \mathbf{b}, \tilde{\beta}, \Lambda, \rho) \propto P(Y|\mathbf{u}_{i}^*, U_{-i},\mathbf{a}, \mathbf{b}, \tilde{\beta}, \Lambda, \rho) \times P(\mathbf{u}_{i}^*).
\end{equation}
$\mathbf{u}_{i^*}$ is independent of all other rows of $U$. The first part requires the evaluation of the bivariate probit likelihood where pairs $(y_{ij},y_{ji})$ are independent of all other pairs of observations. For the second component, recall that a uniform prior is used for $U$, and the new sample is drawn from a symmetric proposal distribution. 
If $u_.$ is not accepted then the state space for the $Z$s does not change and they can be updated conditional on the other $Z$s.

\subsubsection{Other parameters}
The updates for the other parameters, $\Sigma_{ab}$ and $\rho$ proceed in the standard way: $\Sigma_{ab}$ leverages the conjugacy of the Inverse-Wishart distribution, while $\rho$ is updated via a Metropolis Hastings step (that is conditioned on the current value of $Z$). 

\subsection{Posterior inference and summarization}\label{posteriorInference}
If there is no prior information about the communities and a uniform (symmetric) prior is placed on the community labels captured by $U$ then the posterior will also be symmetric \citep{stephens2000dealing}. This is problematic since this leads to multiple posterior modes, all with equal probability, and thus potential label switching as different modes are explored. In our setup, this means that we should consider inference on $U\tilde{\boldsymbol\beta}$ rather than $\tilde{\boldsymbol\beta}$ directly. We thus use a post-processing procedure suggested in \citet{stephens2000dealing} which has become a popular way for dealing with label switching \citep{krivitsky2009representing,hurn2003estimating,fraley2002model}. This is done by first obtaining the $95\%$ posterior credible intervals for each individuals' $\mathbf{u}_i \tilde{\boldsymbol\beta}$. Then, to investigate the differences between community coefficients, clustering using k-means is performed on the posterior means of $U\tilde{\boldsymbol\beta}$ to identify which cluster the posterior mean belongs to. Call that cluster $\hat k_i$ for unit $i$. 
Then the posterior credible intervals are calculated for $\tilde{\beta}_k$ 
by averaging the credible intervals of $u_j\tilde{\boldsymbol\beta}$ for units $j$ for which $\hat k_j = k$. 
This step may not be necessary if a non-uniform prior is placed over $U$ as community labels could be identified by community proportions since the posterior would no longer be symmetric.

\subsection{Extending to the Censored Binary Likelihood}\label{ssec:censor}
Many data collection procedures lead to the observed network $Y$ being censored by limiting the number of friends each person can nominate. Denote the maximal number of friends a person can nominate by $m$. 

To accommodate this type of data collection, we introduce a new random effect, $h_i$. This parameter can be interpreted as an offset of the latent variable that measures censoring: for an individual who is not censored, this is set to $0$ while for a censored individual, it is negative. This can be related to standard censored binary regression models when there is a continuous latent outcome. For example, if we set all of the $h_i=h<0$ for censored individuals and $h_i=0$ otherwise, this is equivalent to having a heterogeneous cutoff for censored versus non censored individuals.
To provide more flexibility, we allow $h_i$ to be different for each censored individual. 
Eq ~\eqref{eq:amen_dep} is extended to:
\begin{equation}\label{amen_dep_censored}
Z_{ij} = \beta_0 + \sum_{l=1}^p (\tilde{\boldsymbol\beta}_{rl}\mathbf{u}_i x_{rli} +  \tilde{\boldsymbol\beta}_{cl}\mathbf{v}_jx_{clj}) + \mathbf{u}_i^t\Lambda \mathbf{v}_j + a_i + b_j + h_i 1_{i \in C}  + \epsilon_{ij}
\end{equation}
where a truncated normal prior is placed on the vector of random effects $\mathbf{h}$ such that $\mathbf{h} \sim N(0, \Sigma_h)\mathbbm{1}_{h < 0}$ with $\Sigma_h= \sigma_h^2 I_{n \times n}$ where $\sigma^2_h$ has an inverse-gamma prior. Let $1_C$ be an $n$ dimensional indicator vector that is a $1$ if an individual is censored and $0$ if they are not.  Let $p = 1$ and decorrelate such that:
\begin{center}
      $\tilde{Z} = Z-\beta_0 \mathbf{11}^t - \diag(\tilde{\boldsymbol\beta_r}U^t)X_{r} -X_{c}\diag(\tilde{\boldsymbol\beta_c}V^t)- \mathbf{a1}^t- \mathbf{1b}^t - U \Lambda V^t$ with
 $\tilde{Z}_{*} = s\times \tilde{Z} + t\times \tilde{Z}^t $
 \end{center}
 \begin{center}
    $H_h = s \times (\mathbf{1}_C  \otimes I_{n}) + t \times (I_{n } \otimes \mathbf{1}_C)$.
\end{center}
For censored individuals, the posterior is then:
\begin{center}
$\mathbf{h} \sim N(V_h m_h, V_h^{-1})\times  \mathbbm{1}\{-\infty < \mathbf{h} < 0\}$ where $m_h = vec(\tilde{Z})^t H_h + \mu_h \Sigma_h^{-1} $ and 
$V_h = H_h^t H_h + \Sigma_h^{-1}$.
\end{center}
\section{Simulations and Computational Considerations}\label{simulations}
We demonstrate the performance of our approach on synthetic data. First we compare our approach to some general alternatives including those with oracle knowledge of any underlying groups. Next we describe the initialization approach for the cluster memberships and demonstrate their importance for the mixing of the MC.  We then evaluate the computational burden of the discrete communities, finding that the MH step can in fact increase efficiency in these parts of the chain. Lastly we show the performance of the proposed censored model from Section~\ref{ssec:censor}.
Data are simulated using the model proposed in Section \ref{fullModel} with $K = 3$ equally sized communities. Let $X_r=X_c= X$.  Covariates $X_p$ are drawn from $ N(0, 1)$. Coefficients, $\tilde{\beta}$, are defined in the individual simulations, and the intercept is chosen such that the observed network is approximately $30\%$ dense. Error terms are correlated such that:
 \begin{center}
 $\Sigma_{\epsilon} = 
\begin{bmatrix}
1 & 0.9\\
0.9 & 1
\end{bmatrix}
$ 
\end{center}
and covariance between random effects for an individual $(a_i,b_i)$ is:
 \begin{center}
$\Sigma_{ab} = 
\begin{bmatrix}
1 & 0.5\\
0.5 & 1
\end{bmatrix}
$. 
 \end{center}
\subsection{Binary Likelihood with Community Dependent and Independent Covariates}
Let $n=150$, $\beta_r = 1$, $\beta_c = 2$, $\tilde{\boldsymbol\beta}_r = [1, 0 ,-1]$, and $\tilde{\boldsymbol\beta}_c = [0, -2, 2]$ with $U = V$.
That is, the coefficients are community dependent for $ X_{2,c}$ and $ X_{2,r}$. Note that for both $\tilde{\boldsymbol\beta}_r$ and $\tilde{\boldsymbol\beta}_c$, $\sum_{k = 1}^K \frac{\tilde{\boldsymbol\beta}_k}{K} = 0$. Latent $Z$ is generated as:
\begin{equation*}
Z = \beta_0 \mathbf{11}^t + X_{1r}\beta_r + X_{1c}\beta_c + \diag(\tilde{\boldsymbol\beta}_r U^t)X_{2r} + X_{2c} \diag(\tilde{\boldsymbol\beta}_c U^t) + \mathbf{a 1}^t+\mathbf{1 b}^t + U\Lambda U^t + E.
\end{equation*}
We compare the performance of the models presented in Table \ref{methodstable}. In addition to the community dependent model with varying initialization of communities, we fit standard AMEN models with varying multiplicative effects \citep[\textbf{amen} package,][]{amenPackage}. We also study models on partitioned data where the partition uses either the true or estimated community labels. We show that allowing the model to learn the community labels rather than using fixed estimates yields much more accurate inference. Each model runs for $150,000$ iterations. 
\begin{table}[H]

	\caption{Table of different models considered in simulations}

	\centering
	\begin{tabular}{|c|c|l|l|c|}
		\hline
	ID & joint? & Mult. Eff & Community?  & Comm. coef.\\ \hline
		  {\fontfamily{pcr}\selectfont a} & \cmark & None   			& NA	& \xmark \\
	 	  {\fontfamily{pcr}\selectfont b} & \cmark & $\mathbb{R}^3$ 	& NA	& \xmark \\
	 	  {\fontfamily{pcr}\selectfont c} & \cmark & $\{0,1\}^3$ 		& NA	& \xmark \\
    	  {\fontfamily{pcr}\selectfont d} & \xmark & NA 				& Estimated before	& \cmark \\
	 	  {\fontfamily{pcr}\selectfont e} & \xmark & NA 				& Oracle	& \cmark \\
	 	  {\fontfamily{pcr}\selectfont f} & \cmark & $\{0,1\}^3$ 		& Estimated before	& \cmark \\
    	  {\fontfamily{pcr}\selectfont g} & \cmark & $\{0,1\}^3$  		& Oracle	& \cmark \\
	 	  {\fontfamily{pcr}\selectfont h} & \cmark & $\{0,1\}^3$ 		& Estimated during	& \cmark \\
	 	  {\fontfamily{pcr}\selectfont i} & \cmark & $\{0,1\}^3$  		& Estimated during	& \cmark (all) \\
		\hline
	\end{tabular}
\label{methodstable}
\end{table}

\begin{figure}[H]
	\centering
	\begin{minipage}[t]{\dimexpr.5\columnwidth-1em}
		\centering
		\includegraphics[width=\columnwidth]{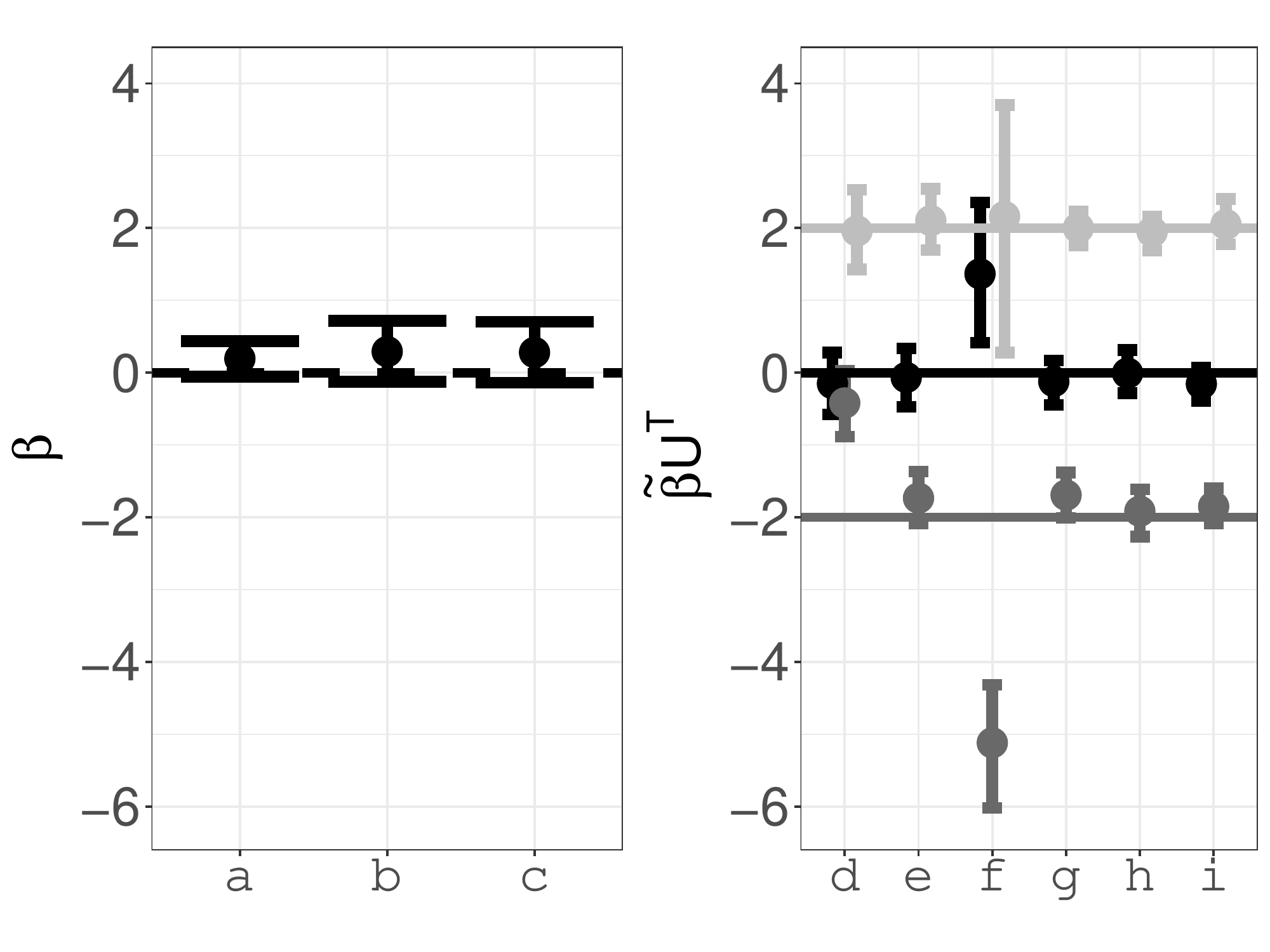}
	\end{minipage}\hfill
	\begin{minipage}[t]{\dimexpr.5\columnwidth-1em}
		\centering
		\includegraphics[width=\columnwidth]{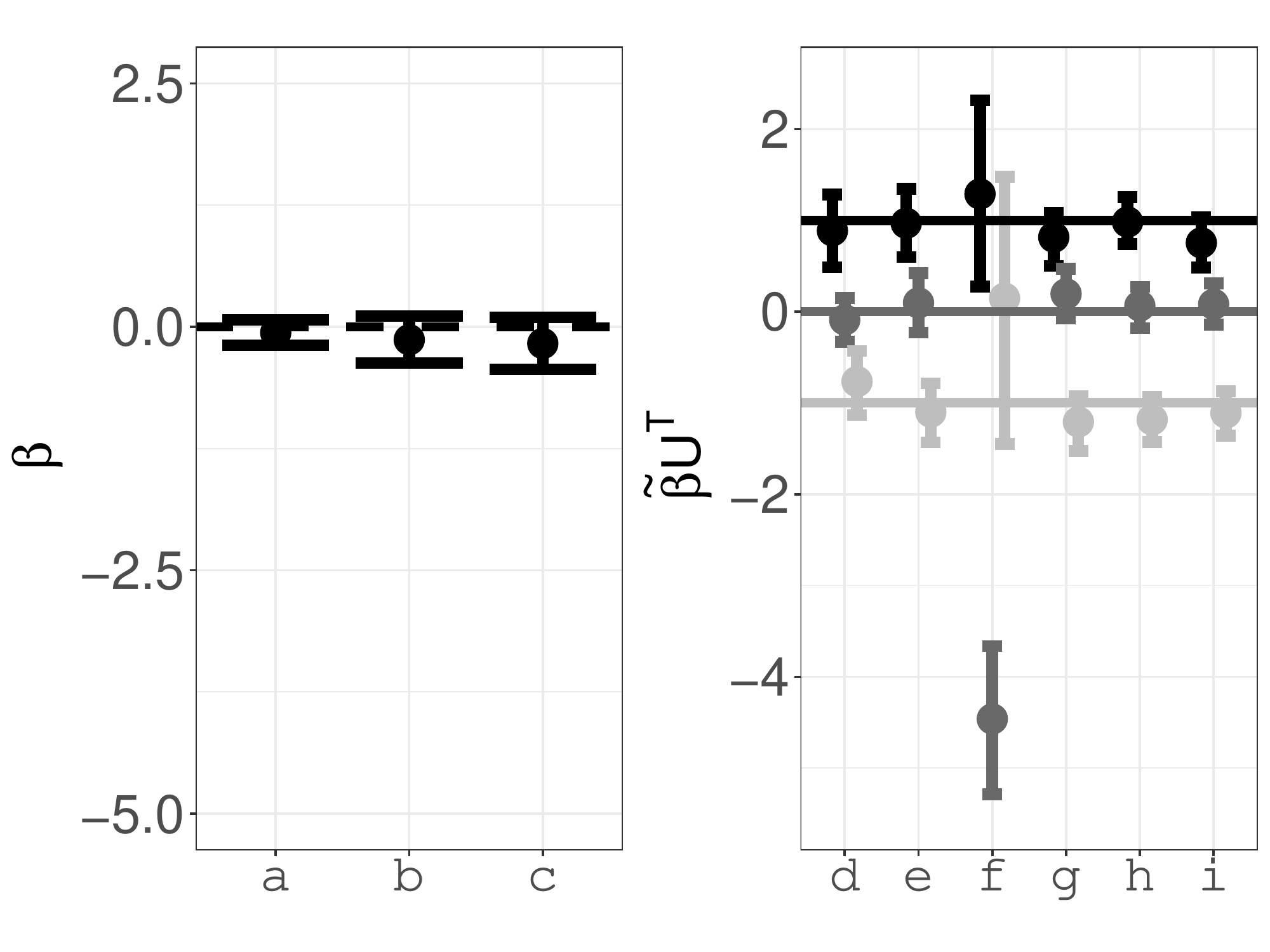}
	\end{minipage}
	\caption{Dashed horizontal line indicates the average $\sum_{k = 1}^K \tilde{\beta}_k/K$ for dependent covariates and solid horizontal lines are equal to the true $\tilde{\beta}$ (Figure on the left are column covariates, figure on the right are row covariates) \label{fig:depRowCol}}
\end{figure}

\begin{figure}[H]
	\centering
	\begin{minipage}[t]{\dimexpr.5\columnwidth-1em}
		\centering
		\includegraphics[width=\columnwidth]{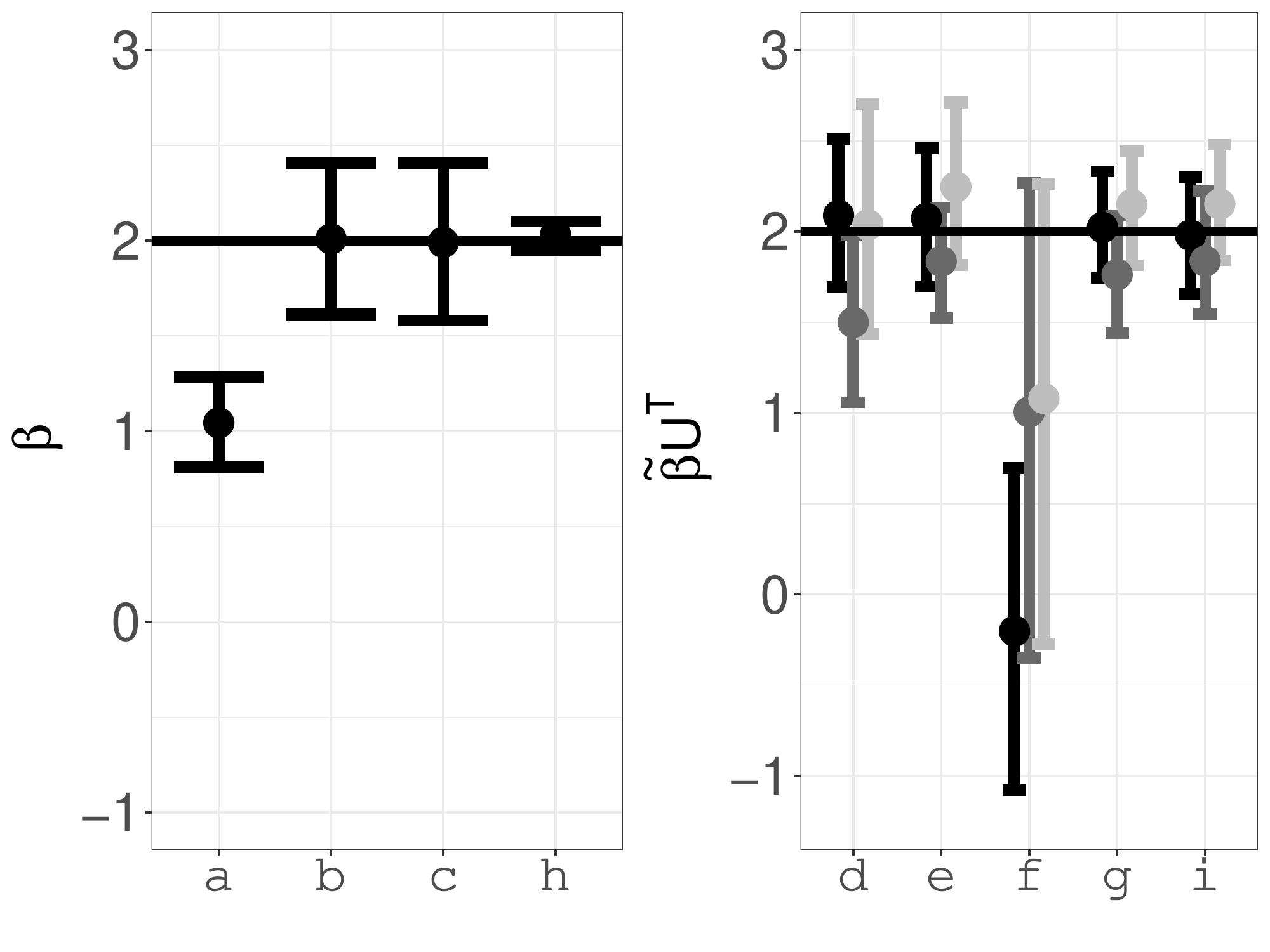}
	\end{minipage}\hfill
	\begin{minipage}[t]{\dimexpr.5\columnwidth-1em}
		\centering
		\includegraphics[width=\columnwidth]{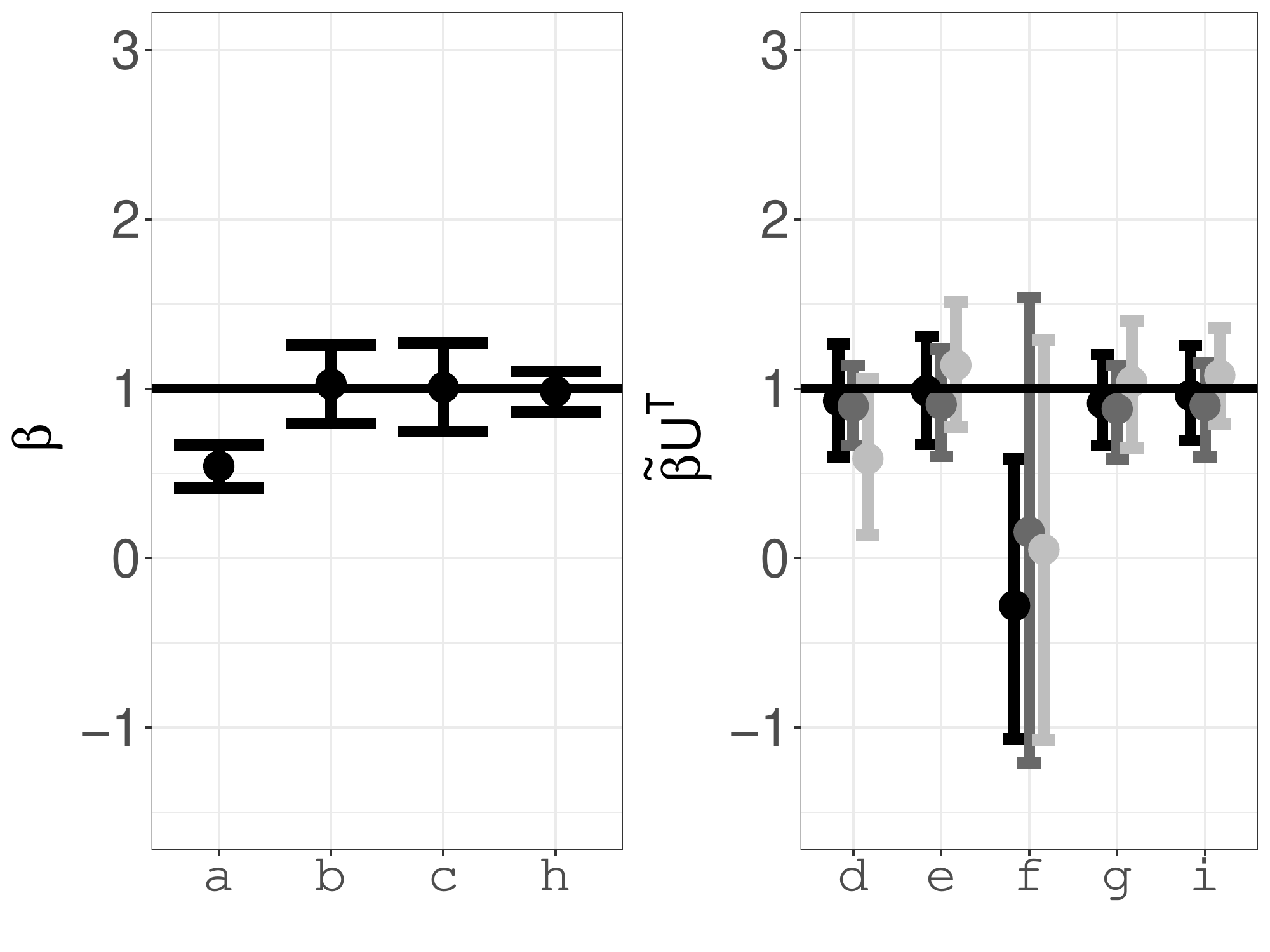}
	\end{minipage}
	\caption{Solid horizontal lines are equal to the true $\beta$ (Figure on the left are column covariates, figure on the right are row covariates) \label{fig:indepRowCol}}
\end{figure}
Figures \ref{fig:depRowCol} and \ref{fig:indepRowCol} plot the 95\% posterior credible intervals for the coefficients for column and row covariates, respectively. The first three models can only estimate community independent coefficients while the remaining models have varying levels of adaptability. We note that model \ml{h}, which is the full model that learns the community membership of units, has oracle knowledge of which covariates are community dependent and hence is grouped with the first three in Figure~\ref{fig:indepRowCol}.
The AMEN model with multiplicative dimension of $3$ (model \ml{b}) yields posterior credible intervals that contain the true value of the coefficients for independent covariates (those associated with $\beta_r$ and $\beta_c$).   The standard AMEN models (\ml{a}, \ml{b}, \ml{c}) yield posterior credible intervals that capture the average of the community dependent $\tilde{\beta}$, exhibiting the loss of information when community dependence is not considered. That is, we might incorrectly conclude that $X_{2r}$ and $X_{2c}$ do not appear to influence connections in $Y$  when indeed two communities are impacted by $X_2$ substantially and one community is not. 
The results for models 	  \ml{d} (separate models for pre-estimated communities) and \ml{f} (full model that uses pre-estimated communities) demonstrate how using the initial estimate of community structure is not sufficient. For both of these, the posterior credible intervals fail to capture the true coefficient values. 
 When $U$ is known, the posterior $\tilde{\beta}$ can be approximated by the data with partitioning as in model 	  \ml{e}, but any cross community information is unknown (e.g. cross community dyadic effects). Nonetheless, this performs nearly as well on both community dependent and community independent covariates as the oracle model that does not partition the data (model 	  \ml{g}). 
 Lastly consider two versions of our proposed model: model 	  \ml{h} assumes prior knowledge of which covariates are community dependent and which are not, while model 	  \ml{i} does not have this knowledge and fits a community dependent coefficient for each covariate. The two models yield very similar results. Importantly, model \ml{i} yields nearly equivalent intervals across communities for $X_1$ while still identifying the varying coefficients associated with $X_2$. This shows how our approach can differentiate between covariates that are not dependent on community structure without being given that prior knowledge.

\subsection{Initializing $U$}\label{initialU}
High dimensional discrete parameters are notoriously hard to work with. However, we saw in model \ml{d} that some communities can be learned prior to model fitting. This suggests that initializing the MCMC at a good estimate of communities might make it easier for the chain to move. Here we explore two approaches to initializing $U$ and show how choosing one of these leads to an improvement in the autocorrelation of the $\mathbf{u_i} \tilde{\beta}$s.
\begin{itemize}
\item \noindent\textbf{Spectral Methods.} Spectral methods leverage an embedding of the adjacency matrix in a Euclidean space and then estimate communities by clustering in the embedded space. There are many variations on spectral methods \citep{Rohe11,rohe2017covarAssist,abbe:2018, CAMSAP2019, Suwan16}, but here we concentrate on a normalized graph Laplacian embedding. Specifically, let
	$L = D_{\tau,c}^{-1/2}YD_{\tau,r}^{-1/2}$ be the normalized Laplacian where 
	$D_{\tau, r} = D_r + \tau_r I_n$, $D_{\tau, c} = D_c+ \tau_c I_n$ (these are adjusted diagonal matrices that capture the degree of each node: 
 $D_{r,{ii}} = \sum_{j =1}^n Y_{ij}$, $\tau_r = \frac{\sum_{i=1}^n D_{r,{ii}}}{n}$ and $D_{ c,{ii}} = \sum_{i =1}^n Y_{ij}$, $\tau_c = \frac{\sum_{i=1}^n D_{c,{ii}}}{n}$).
Compute the $K$ left singular vectors of $L$ and cluster them using $k$-means clustering. Initialize $U$ to be the matrix corresponding to those clusters.

\item \noindent\textbf{Residual Clustering.} 
An alternative approach is to initialize $U$ using residuals obtained from performing probit regression (this is in effect a feasible GLS estimate of $U$). The procedure is as follows: fit a simple probit regression model to the observed data,
\begin{center}
	$P(vec(Y) = 1|X) = \Phi(vec(X) \beta)$,
\end{center}
and write the residuals in matrix form. Take the singular value decomposition of this residual matrix and apply clustering methods on the top $K$ left singular vectors, similar to in the previous method.
\end{itemize}
We note that this is simply an initialization for the communities rather than a true estimate for them. In particular, both methods can actually be used in the more general setting where individuals can belong to different row and column communities (this can be done by simply clustering the left and right singular vectors in the above procedures).
\subsubsection{Comparing Initializations}\label{initialCompare}
\begin{figure}[H]
 \centering
		\includegraphics[width=.8\columnwidth]{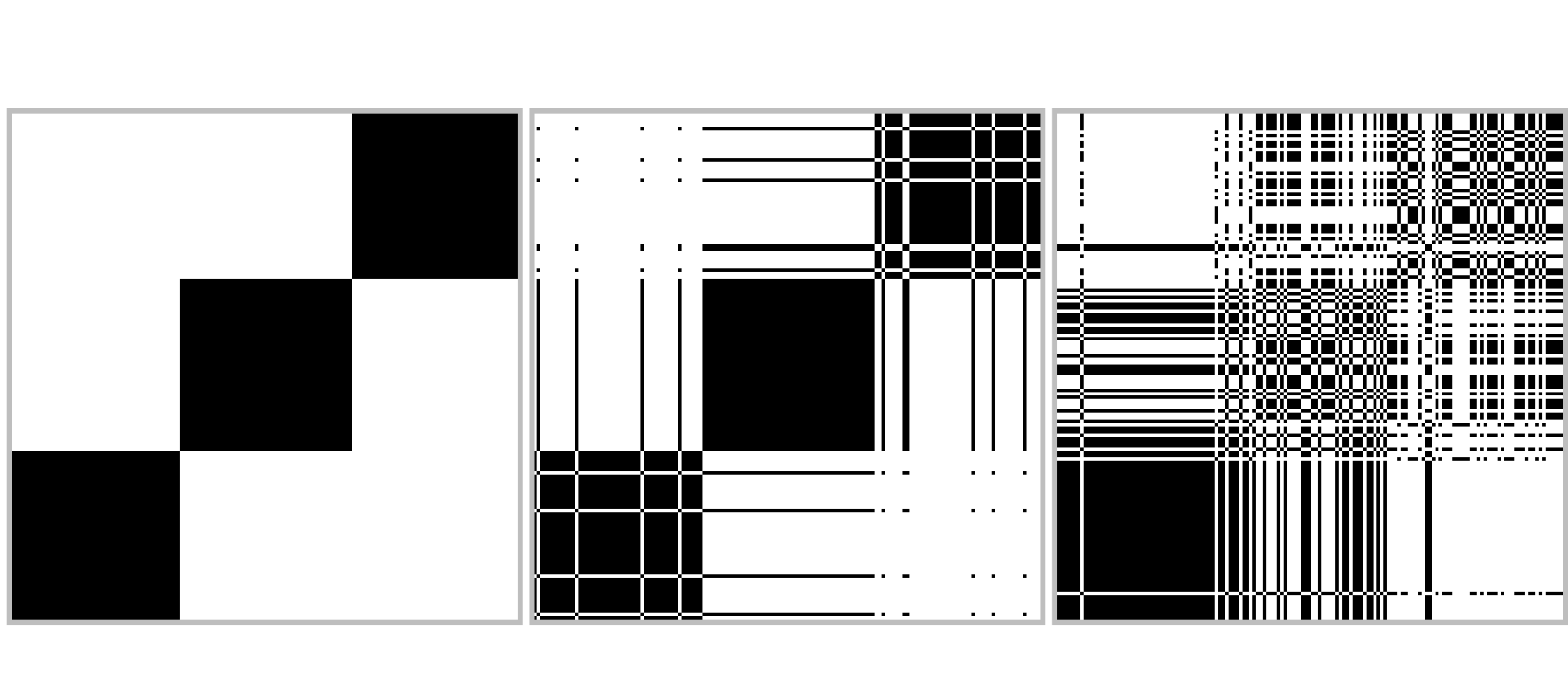}
		\caption{From left to right, true $UU^t$ from the model , $\hat{U}\hat{U}^t$ where $\hat{U}$ contains membership estimates using non covariate assisted spectral clustering, $\hat{U}\hat{U}^t$ where $\hat{U}$ contains member estimates from clustering on probit regression residuals. \label{fig:initialCompare}}
\end{figure}

For a directed network that we study in detail in Section \ref{simulations} we plot the true community structure (left panel of Figure~\ref{fig:initialCompare}), the estimated structure using a spectral approach (middle panel of Figure~\ref{fig:initialCompare}) and the residual clustering (right panel of Figure~\ref{fig:initialCompare}). In this case it appears that the spectral method yields an estimate that is slightly closer to the true community structure than the residual clustering approach. We note that in applications there is no direct access to the ground truth and so the inherent nonidentifiability of the order of individuals in a network means that such visualizations require an additional post-processing step to provide a visual indication of community structure \citep{chan2014consistent}.  Nonetheless, visualizations such as this can help through the analysis pipeline by identifying mode changes or mixing properties of the chain. 

\subsubsection{Comparing Convergence from Different Initialization}
\begin{figure}[H]
		\centering
		\includegraphics[width=.5\columnwidth]{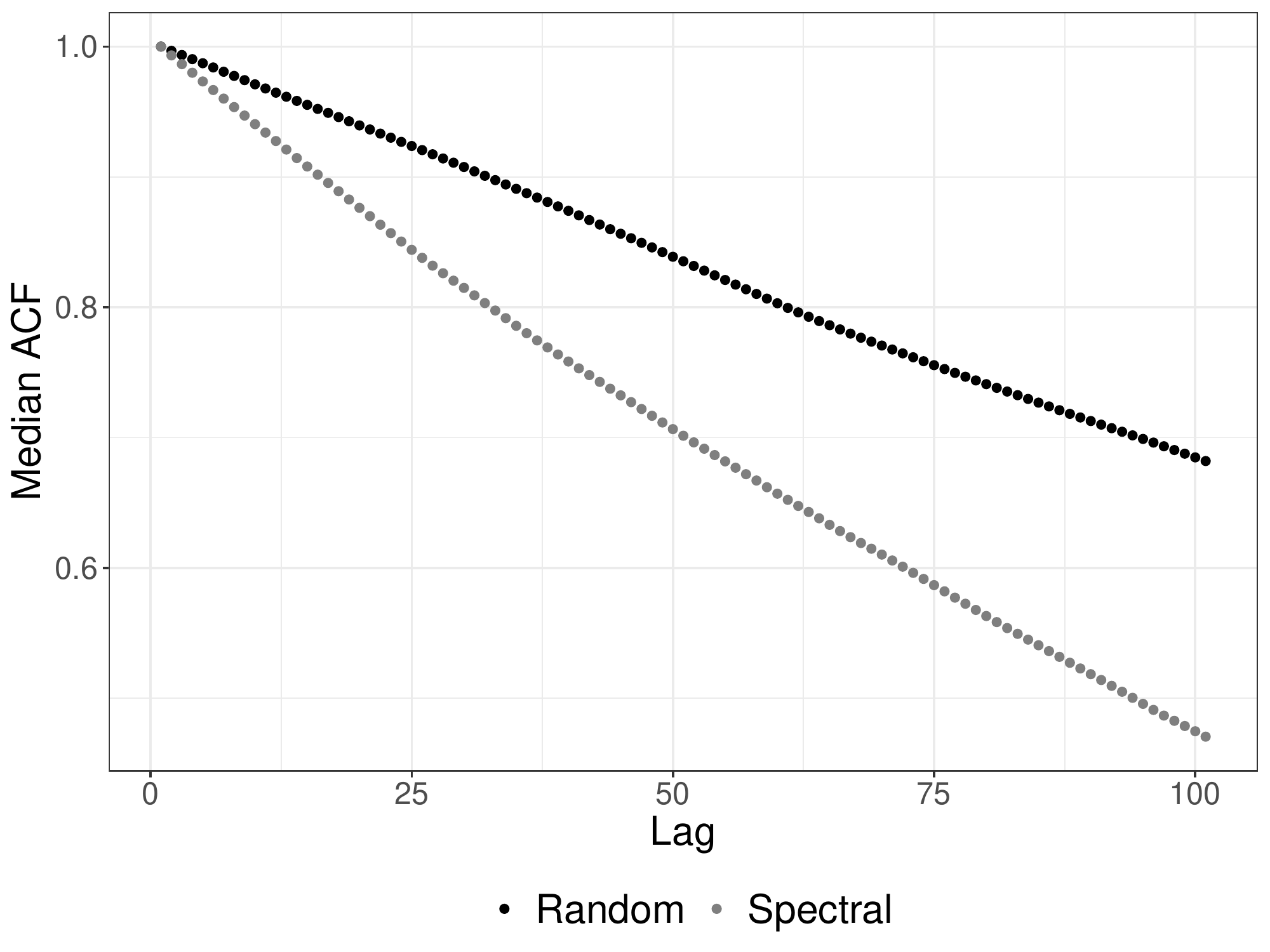}
		\caption{Median ACF for $\mathbf{u}_i \tilde{\beta}$ across $i \in \{1,...,150\}$. \label{fig:acf}}
\end{figure}

Here we evaluate the effect of initialization of $U$ on the induced autocorrelation on $U\tilde{\boldsymbol\beta}$ in the simulations from Section \ref{simulations}.  Specifically we study two runs of the Markov Chain on the same data set, one initializing $U$ uniformly at random and one with the spectral clustering proposed in Section \ref{initialCompare}. Both chains are run for $150,000$ iterations and Figure \ref{fig:acf} provides summaries of the autocorrelations for the $\mathbf{u}_i \tilde{\boldsymbol\beta}$'s at different lags (i.e. each plot summarizes the estimates of the lag$-t$ autocorrelation across all $150$ units).  We see a substantial difference in the rate of decay of the ACF with the spectral decomposition exhibiting less autocorrelation and lower lags than the uniform initialization. Similar behavior is observed for other functions of the posterior.

\subsubsection{Computational Advantages}\label{compAdvantage}
One of the important advantages of the latent space model of \citet{hoff2002latent} is its' flexibility in capturing different nodal behaviors \citep{hoff2008modeling}. In this section we address whether restricting the latent multiplicative positions to be discrete introduces a trade-off between computation and accuracy of estimates. We generate data from Eq ~\eqref{eq:OGamen}, incorporating two non-community-dependent covariates and setting the latent variables to be independent samples from a 3-dimensional normal distribution $\mathbf{u_i}\sim  N_{3}(0, I_{3})$. We fit the correctly specified AMEN model, an under specified AMEN model without multiplicative effects, and our proposed model with 3 communities and the proposed update schema. 
Performance is measured using effective sample size divided by a measure of accuracy for each coefficient, $\beta_{err} = (\hat{\beta} - \beta_{true})^2$ where $\hat{\beta}$ indicates the estimated posterior mean. This allows us to visualize the trade-off between speed and estimation quality. Note that higher $ESS/\beta_{err}$ is desirable. 
\begin{figure}[H]
	\centering
	\includegraphics[width=0.6\columnwidth]{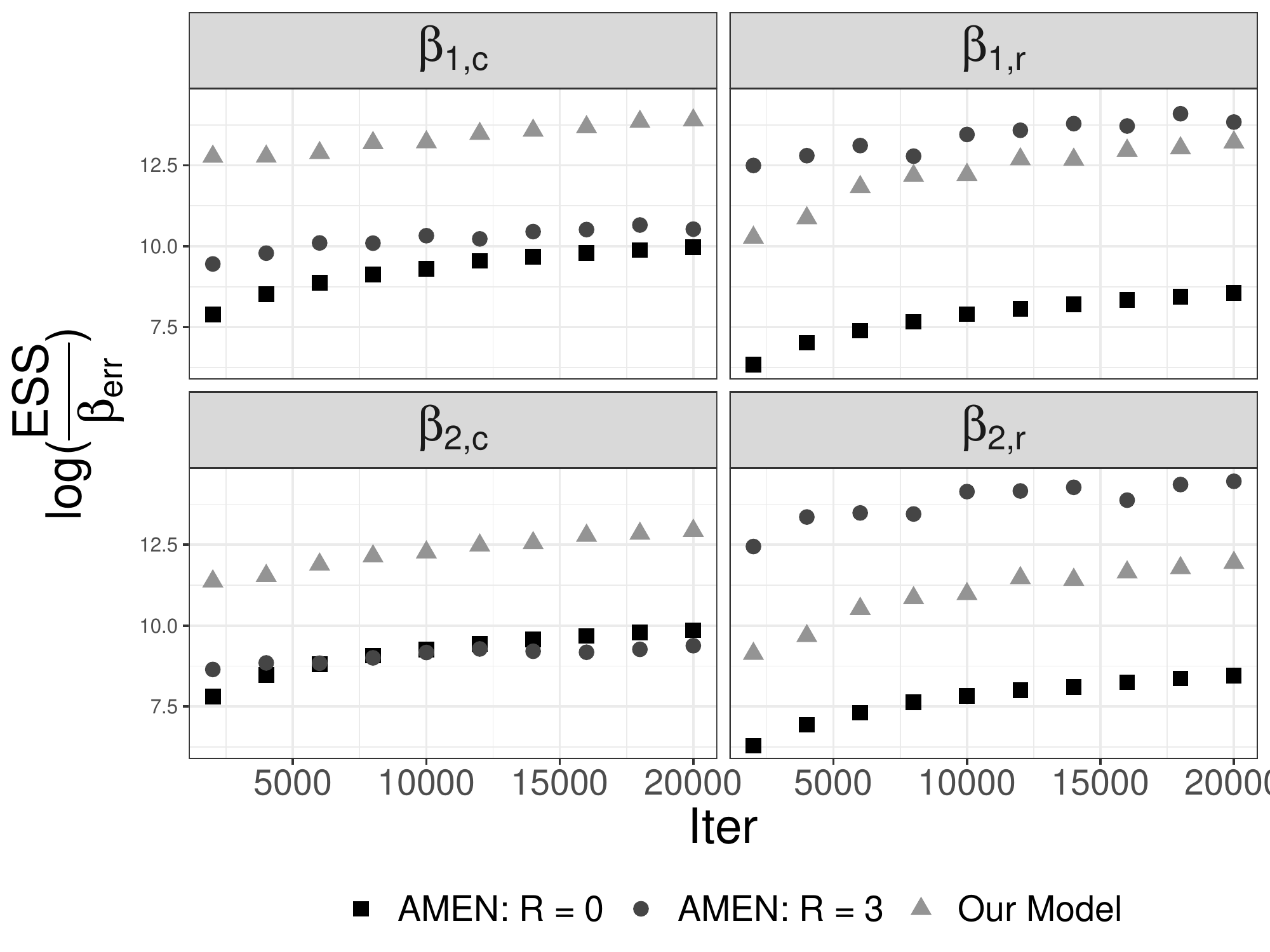}
	\caption{Plots showing number of iterations vs log of effective sample size (ESS) divided by $\beta_{err}$. Each point is an average over $10$ runs. There were $10$ different iteration levels.} 
	\label{TimerFig_bss}
\end{figure}
Figure \ref{TimerFig_bss} plots these tradeoffs for the three models. We see that AMEN with no multiplicative effects performs the worst as expected since it is under specified and unable to account for excess variance. Our model, on the other hand, performs on par with the correctly specified AMEN model, even slightly outperforming it for two of the parameters. The move from continuous multiplicative effects to binary ones affords a computational improvement (especially with the marginal updates we propose). This figure showcases that this improvement comes with little loss in accuracy in estimating parameters even when the multiplicative effects are misspecified.

\subsection{Effect of Censoring}\label{ssec: censored_simul}
\begin{figure}[H]
	\centering
	\includegraphics[width=.45\columnwidth]{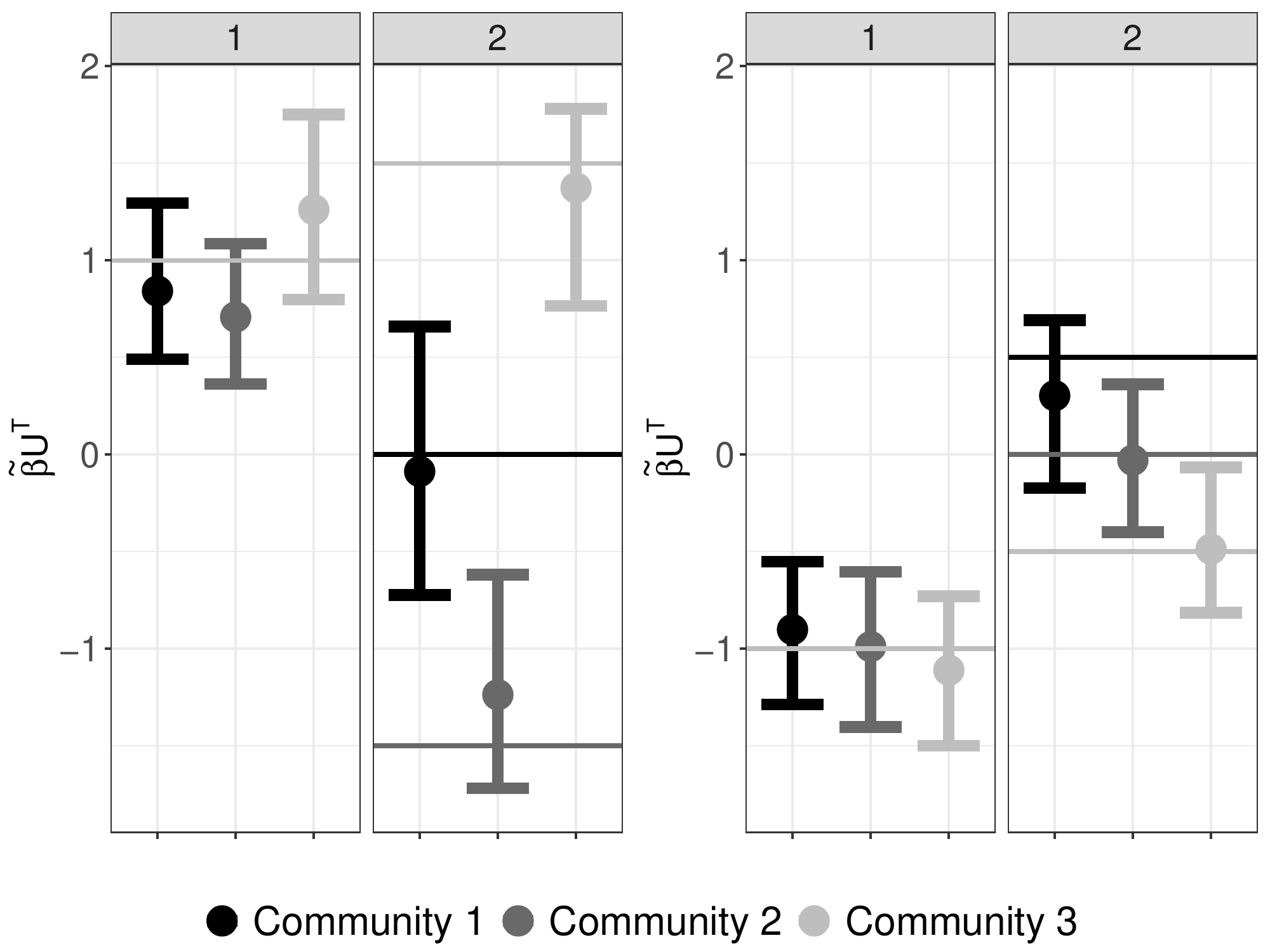}
	\caption{$95\%$ Credible intervals for row and column covariates from censored community dependent model \label{fig:censoredPosterior}}
\end{figure}

Lastly, we demonstrate the performance of the proposed model when censoring of data occurs (as it does for the AddHealth data we analyze in Section \ref{realData}). Let $n =150$ with $K = 3$ equally sized communities. 
The covariates are distributed as in Section~\ref{simulations} with coefficients given by
\begin{center}
$\tilde{\beta}_{c} = 
\begin{bmatrix}
1 & 1 & 1\\0 & -1.5 & 1.5
\end{bmatrix}$ ,
 $\tilde{\beta}_{r} =\begin{bmatrix}
-1 & -1 & -1\\0.5& 0 &- 0.5
\end{bmatrix} $
\end{center} 
where individuals are allowed to have a maximum of $15$ friends (this creates a network where about about $10\%$ of connections are censored). We fit model~\ml{i}, which includes community-based coefficients for each covariate, and Fig~\ref{fig:censoredPosterior} plots 95\% credible intervals for each of the coefficients. We see that while censoring increases the uncertainty about the estimates, the posterior is still able to capture the mean behavior, even in the presence of censoring. Supplement S5 showcases the similarity in inference between the specification of the censored binary likelihood in \cite{hoff2013amen} and in Section~\ref{ssec:censor}. 

\section{Application to AddHealth Data}\label{realData}
We come back to study the friendship formation behavior of American high school students. These data were collected as part of the National Longitudinal Study of Adolescent Health (AddHealth) dataset \citep{addHealth2009}. This nationally representative study of adolescents during the 1994-1995 school year includes survey information on school and home activities and behaviors as well as information on the top five male and top five female friends for each student.
Previous literature has demonstrated the importance of individual and pairwise measurements of smoking and drinking behaviors, race and extracurricular involvement on same-sex friendship formation \citep{hoff2013amen}. It has further been postulated that home composition (such as living with one's mother or father) can influence in-school behavior and so we include these indicators in our analysis \citep{dunn2001family}. All covariates (described in detail in Supplement S6) are incorporated into our analysis as potentially community informed sender, receiver and dyadic effects.
Missing covariate values are imputed before the analysis using a semi-parametric Gaussian copula approach  \citep{sbgcop2007, sbgcop,hollenbach2018multiple}.

We illustrate the flexibility of our modeling framework by studying the male-male networks of two different schools. School A is a public, rural high school in the American South with 205 male students while School B is a public, suburban school in the American Northeast with $292$ males. Both schools are majority white --- $80\%$ and $92\%$ respectively --- and approximately the same fraction of each school nominated the maximal number of friends ($86$ and $123$, respectively).

\begin{table}[H]
	\centering
	\begin{tabular}{l||cc||cc}
		& \multicolumn{2}{c||}{School A} & \multicolumn{2}{c}{School B} \\
		& Community 1 & Community 2 		& Community 1 & Community 2 \\ \hline
		Smoke           & 33.94\%   & 23.96\% & 38.30\% & 17.88\%   \\
		Drink           & 33.94\% & 25.00\% & 36.17\% & 37.75\% \\
		Smoke and Drink & 22.02 \% & 12.5\% & 25.53\% & 15.89\% \\
		Live with Mom   & 90.08\% & 85.42\% & 95.74\% & 94.03\% \\
		Live with Dad   & 77.98\% & 79.17\%   & 91.49\% & 84.77\% \\
		White           & 73.39\%   & 86.46\% & 95.75\% & 89.40\%\\
	\end{tabular}
\caption{Comparing percent of students with a particular response by estimated latent community}
\label{tab:covByComm}
\end{table}

\subsection{School A} 
\begin{figure}[H]
	\centering		\includegraphics[width=.7\columnwidth]{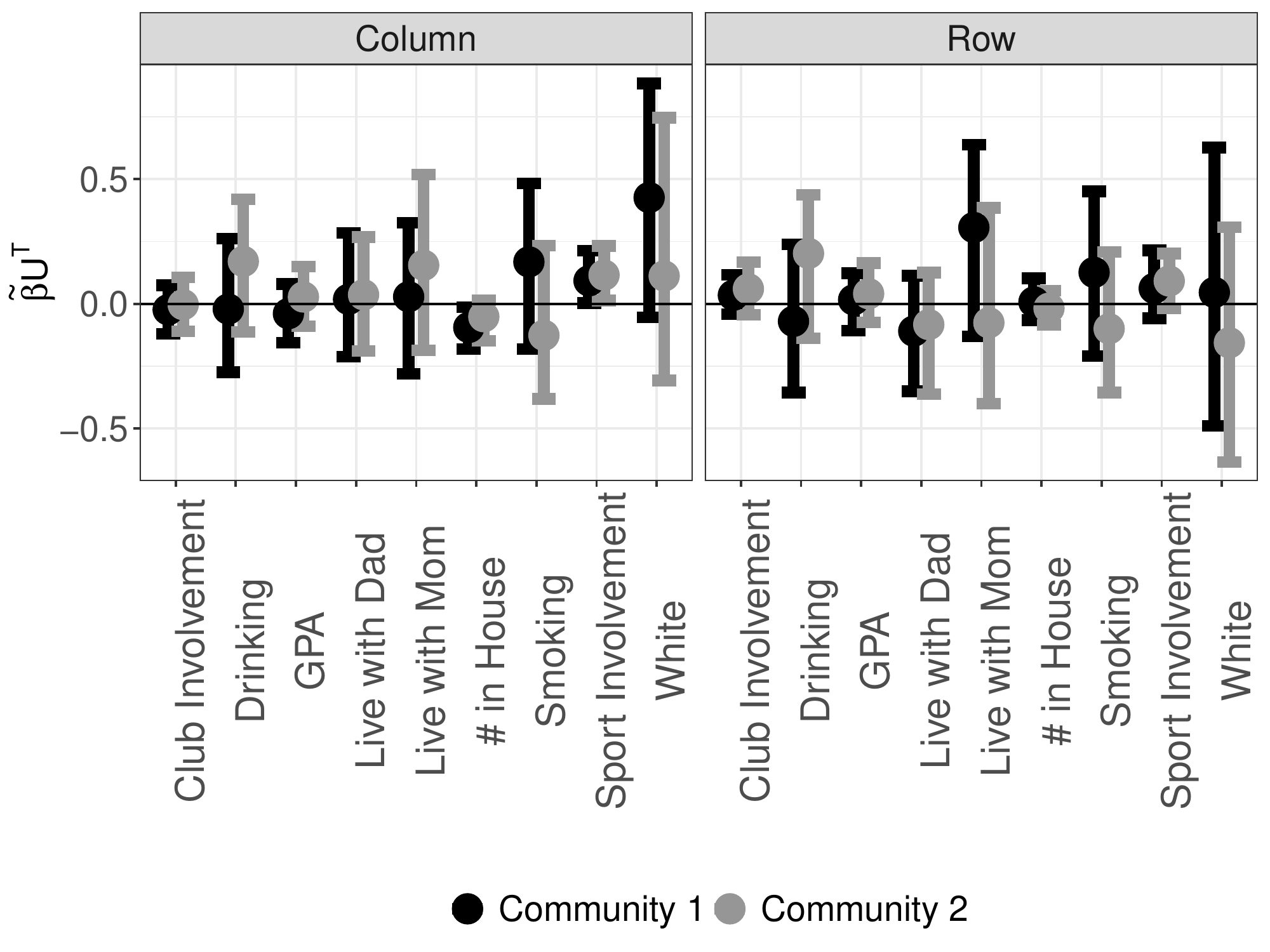}
	\caption{$95\%$ CI for row and column covariates with community dependent model, $K = 2$ \label{fig:addHealth22}}
\end{figure}

We first present results for a school where no community structure appears to moderate the effect of covariates on friendship formation. That is, while communities are identified, community dependent coefficients are estimated to be nearly identical to community independent ones. 
Figure~\ref{fig:addHealth22} presents the community-dependent coefficients for both row- and column-effects. While some variability is present, the credible intervals for the community based coefficients are largely overlapping and the conclusions for the two communities are nearly indistinguishable. We see that those who are involved in sports are more likely to send and receive connections, while living in a larger household  appears to reduce ones popularity. Dyadic covariate results are presented in Supplement S8 and showcase that being in the same grade increases the probability of connection regardless of community membership. The only differences we observe between communities relate to the effect of race on friendship formation. While the covariate distribution matches across communities (as demonstrated in the first two columns of Table~\ref{tab:covByComm}), the largest difference between them is in their racial composition. This could contribute to this slight difference in the effect of race on friendship formation. 

\subsection{School B}
The students in this school split into two communities, where each community exhibits substantively different friendship-formation behavior. The communities have $141$ and $151$ students respectively, and importantly (as with School A), the community structure appears latent and cannot be predicted by any combination of covariates (see last two columns of Table~\ref{tab:covByComm}).
Because the school is predominantly white, race is not included in the analysis (this is the only deviation from the model formulation between the two schools).

\begin{figure}[H]
	\centering
	\includegraphics[width=.7\columnwidth]{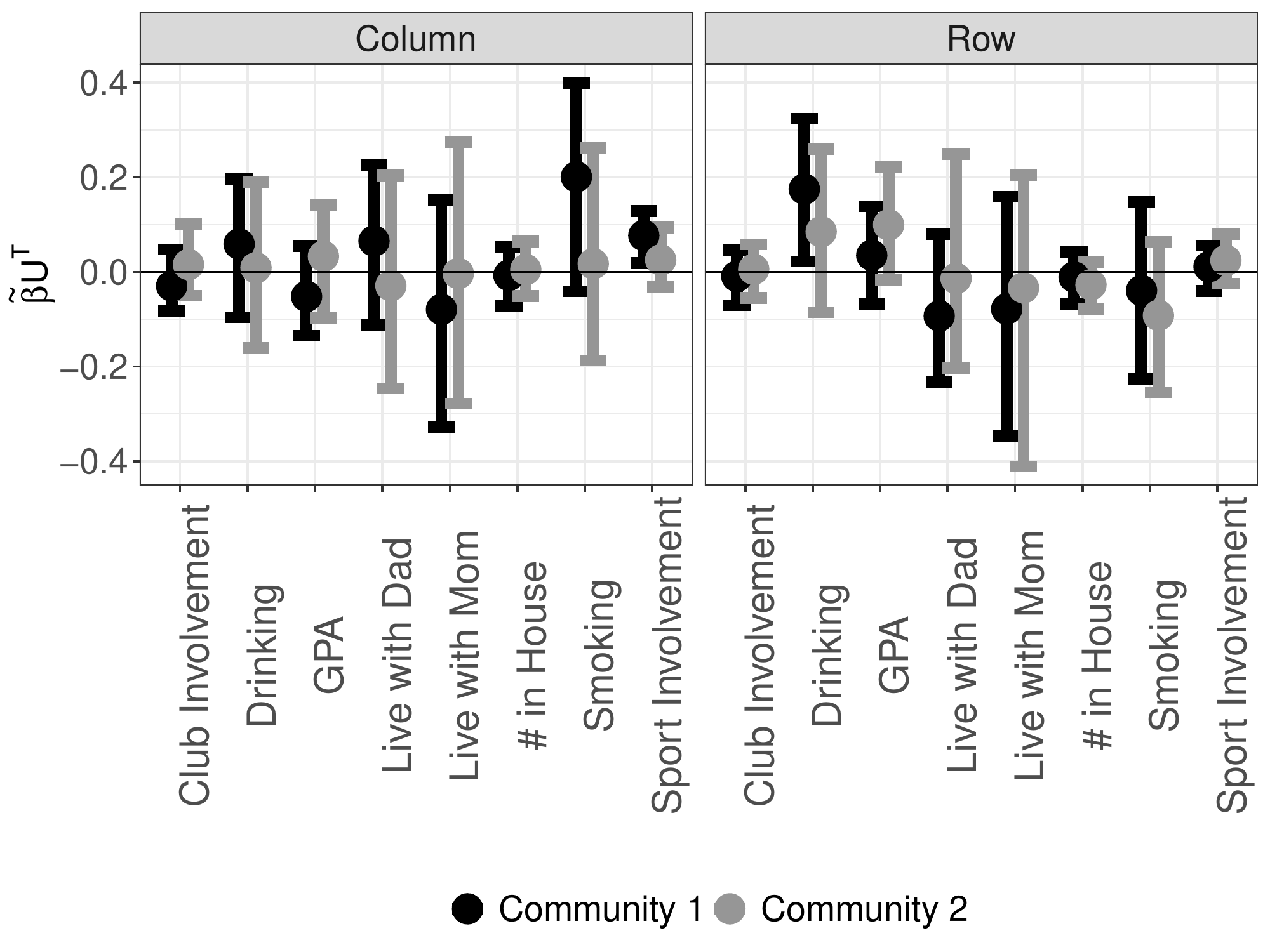}
	\caption{$95\%$ CI for row and column covariates with community dependent model, $K = 2$ \label{fig:addHealth23}}
\end{figure}

Figure~\ref{fig:addHealth23} presents the 95\% posterior credible intervals for the community dependent row and column coefficients (plots for dyadic covariates are in Supplement S8). We see immediate differences between the two communities: in community 1 smoking and sport involvement increase popularity and drinking increases sociability, but these effects are not present for community 2. On the other hand, GPA appears to increase sociability, but only in community 2. While being in the same grade increases the probability of friendship for both communities, the effect is stronger for community 1. Lastly, sharing drinking and smoking behavior increases the probability of friendship for community 1 but not for community 2.

\begin{figure}[H]
	\centering
	\includegraphics[width=.7\columnwidth]{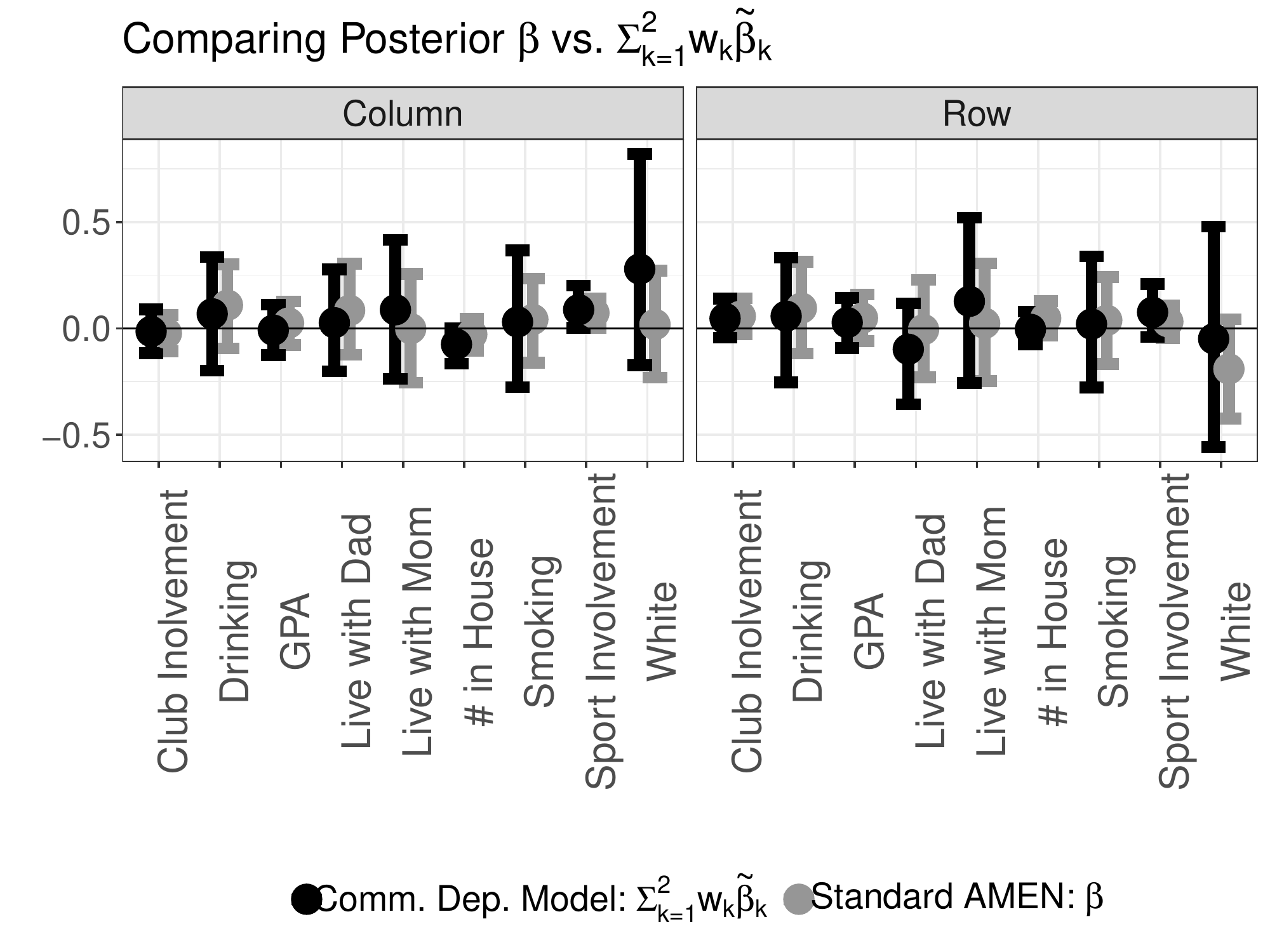}
	\caption{School A: Comparing $95\%$ CI $\sum_{k = 1}^2 \frac{\tilde{\beta}_{k}}{2}$  vs. AMEN with multiplicative effects of dimension 2 for row and column covariates  \label{fig:addHealthCompareModels22}}
\end{figure}

\begin{figure}[H]
	\centering
\includegraphics[width=.7\columnwidth]{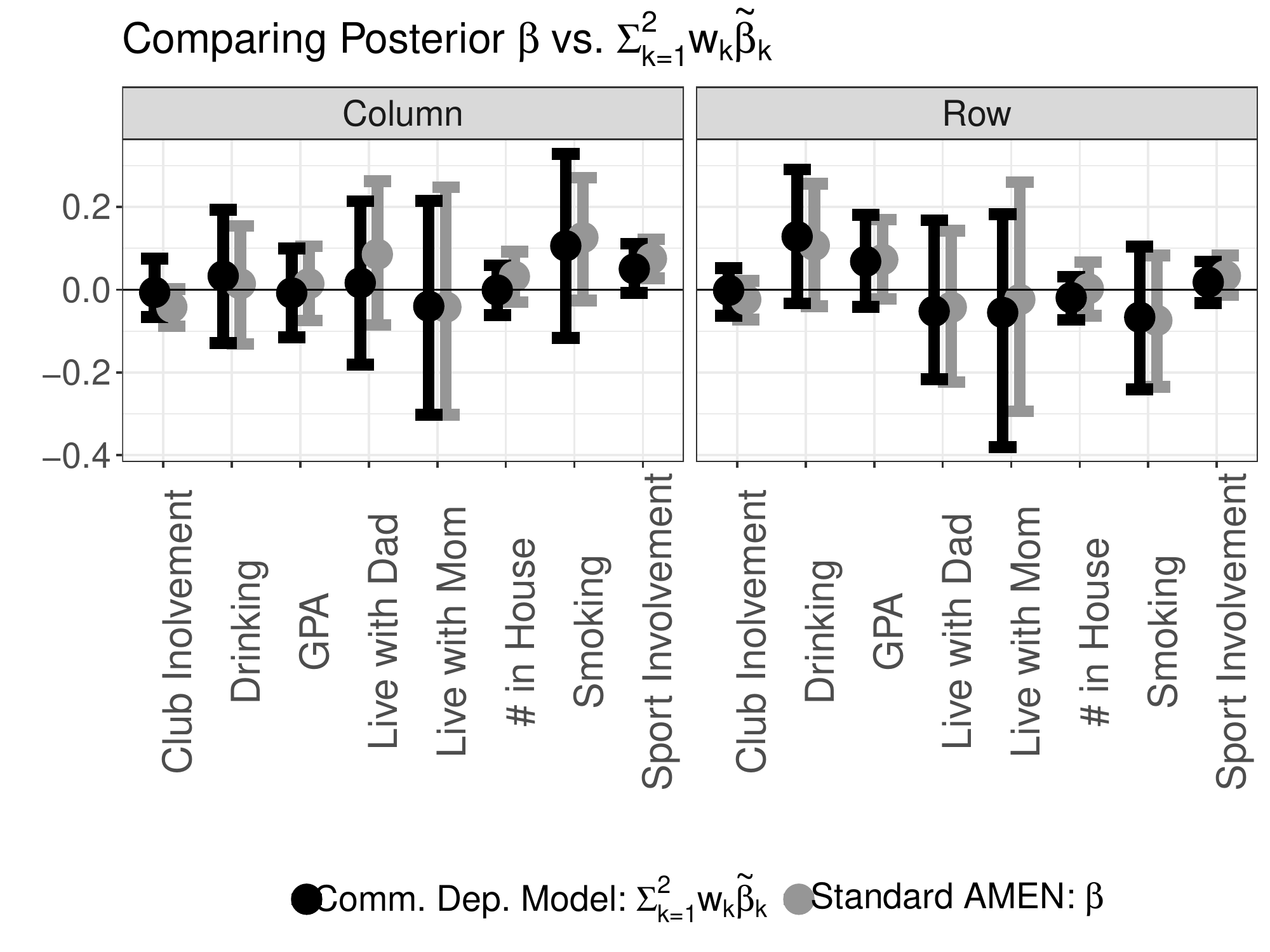}
	\caption{School B: Comparing $95\%$ CI  $\sum_{k = 1}^2 \frac{\tilde{\beta}_{k}}{2}$ vs. AMEN with multiplicative effects of dimension 2 for row and column covariates  \label{fig:addHealthCompareModels23}}
\end{figure}

\paragraph*{Comparison with community independent model} 
Figures \ref{fig:addHealthCompareModels22} and \ref{fig:addHealthCompareModels23} summarize the similarity between the average community weighted coefficients and the output of the less flexible standard AMEN model with 2-dimensional multiplicative effects for Schools A and B, respectively. We see that generally the two models produce similar results when averaging over the communities. This is consistent with the theoretical claims made in Section~\ref{asymptoticB} and the simulations in Section~\ref{simulations}. Importantly, for networks where no community dependence is present, this suggests that estimating the more complex model does not have to come at the expense of high quality average inference. 

However, when community structure is present, the average effect can be zero when the effect is non-zero for one or both communities. For example, in School A we observe a negative effect of household size on sociability but this effect is not present in either model when averaging over the communities. Similarly, smoking appears to increase sociability and drinking appears to increase popularity in the averaged results for School B (Fig.~\ref{fig:addHealthCompareModels23}, but this effect is actually only present in one of the two communities. These discrepancies can lead to significantly different policies. For example, students from larger families might need additional support for integrating into the school community, but this would not be observable if no community dependent model were available. Similarly, while drug and alcohol programs have been demonstrated to be effective in American high schools \citep{eisenberg2003evaluating}, the community dependent model suggests which communities coalesce around these behaviors, which might allow for more targeted interventions.    

\paragraph*{Convergence diagnostics}
As the proposed model includes a large number of parameters, convergence diagnostics must be presented to evaluate whether the Markov Chain Monte Carlo algorithm has run sufficiently long. Visual inspection of traceplots for several parameters (see plots in Supplement S7) suggests that the MCMC is stable. Further, for each $\mathbf{u}_j\tilde{\beta}$ we computed Geweke $z$-scores, with $70\%$ and $92\%$ of those being under 2 across all covariates and individuals for Schools A and B, respectively. We note that running the chains substantially longer would likely lead to more of the Geweke statistics to be below 2, but it is unlikely to change the estimates of the posterior. This conclusion can be evaluated heuristically: for School A there is little community dependence and substantially improved mixing in the community independent model is observed with $\sim \%90$ of Geweke $z$-socres under 2 (which provides estimates that are approximately the same as the community dependent model). This observation suggests that improvements to mixing can be achieved by reducing community dependence where we do not see evidence for it. 
\section{Discussion}\label{discussion}
We have illustrated the importance of accounting for latent community structure when performing inference to determine influence of covariates on forming connections in a network.  To accommodate for this, we have developed a model that efficiently incorporates community structure into the standard AMEN model by using the simultaneous estimation of memberships to allow for community dependent covariate coefficient estimation. This model provides interpretable and meaningful multiplicative effects as well as computationally efficient methods for estimating them. Additionally, the value of this model is exhibited in a real world example where accounting for community dependence allows for better understanding of how networks form in American high schools. 

Possible extensions for our model include allowing for $U$ and $V$ to be continuous for a more flexible model. For example, rather than binary values, each row of $U$ could be defined as a vector of membership probabilities. Then, the resulting value of $\tilde{\boldsymbol\beta}U^t$ would be a mixture of $\tilde{\beta}$. While more flexible, we lose the interpretability that is given with the binary form. Another extension would allow for time dependence in the community structure \citep[which has been demonstrated in some social networks, e.g.][]{xin2017continuous} as well as the covariate coefficients. This in turn introduces another dependence structure that can be challenging to account for. Finally, while not heavily explored in this paper, this model can be used for prediction. Often we are interested in possible future connections between individuals. Given the Bayesian framework used, predictive posterior distributions are easily obtained and predicted network structure can be studied. 

\section*{Acknowledgements}
The first author was partially supported by ARI W911NF1810233. 

The second author was partially supported by NSF CAREER DMS-2046880, ARI  W911NF1810233, and NIH  1R01EB025021. 
\newpage

\bibliography{citations}

\bibliographystyle{apalike}

\newpage

   \begin{center}
      \Large\textbf{Supplement}
   \end{center}

\setcounter{section}{0}

\section{Vectorization of $\diag(\tilde{\boldsymbol\beta}_r U^t)X_r$}

For deriving full conditionals needed to sample our parameters of interest, it is convenient to rewrite  $\diag(\tilde{\boldsymbol\beta}_r U^t)X_r$ in vectorized form (for simplicity we consider $p$ = 1)
\begin{align*}
vec(\diag(\tilde{\boldsymbol\beta}_r U^t)X_r)  &= vec(\diag( U \tilde{\boldsymbol\beta}_r^t)X_r)\\
&= vec((I_n \circ U \tilde{\boldsymbol\beta}_r ^t \mathbf{1}_n^t)X_r)\\
&= vec([I_n \circ X_r] U \tilde{\boldsymbol\beta}_r^t \mathbf{1}_n^t)\\
&= (\mathbf{1}_n \otimes [I_n \circ X_r] U) vec(\tilde{\boldsymbol\beta}_r).
\end{align*}
The main quantities of importance are:
\begin{center}
	$\diag(U \tilde{\boldsymbol\beta}_r ^t) = (I_n \circ U \tilde{\boldsymbol\beta}_r ^t \mathbf{1}_n^t)$
\end{center}
and
\begin{center}
	$(I_n \circ  U \tilde{\boldsymbol\beta}_r \mathbf{1}_n^t) X_r =(I_n \circ X_r)U \tilde{\boldsymbol\beta}_r \mathbf{1}_n^t $
\end{center}
which hold due to the structure of $ U \tilde{\boldsymbol\beta}_r^t \mathbf{1}_n^t$. 

Note that the above is specifically for row covariates. A similar calculation is available for column covariates:
\begin{center}
	$\diag(V \tilde{\boldsymbol\beta}_c ^t) = (I_n \circ V \tilde{\boldsymbol\beta}_c ^t \mathbf{1}_n^t)$
\end{center}
\begin{center}
	$X_c(I_n \circ  V \tilde{\boldsymbol\beta}_c\mathbf{1}_n^t)  = (V\tilde{\boldsymbol\beta}_c \mathbf{1}_n^t)^t(I_n \circ X_c)  $,
\end{center}
\begin{align*}
vec(X_c \diag(\tilde{\boldsymbol\beta}_cV^t))  &= vec(X_c (I_n \circ V \tilde{\boldsymbol\beta}_c ^t \mathbf{1}_n^t))\\
&= vec((V\tilde{\boldsymbol\beta}_c \mathbf{1}_n^t)^t(I_n \circ X_c)  )\\
&= vec(\mathbf{1}_n \tilde{\boldsymbol\beta}_c^t V^t [I_n \circ X_c] )\\
&= ([I_n \circ X_c] V \otimes \mathbf{1}_n)vec(\tilde{\boldsymbol\beta}_c).
\end{align*}
\section{Including  Dyadic Covariates}\label{covarDyad}
Often, dyadic covariates are observed or created from row/column covariates. A dyadic covariate, $X_d$, is a $n \times n$ matrix where each entry $x_{d,{ij}}$ describes some notion of how $i$ relates to $j$. For example, observe the row/column covariate, grade.  A dyadic covariate can be created for grade such that $X_{d,{ij}}= 1$ if $i$ and $j$ share the same grade, otherwise it is $0$. 

Incorporating dyadic covariates in the standard AMEN model is simple but is more complicated when coefficients can be community dependent. To begin, consider a dyadic covariate that has a special form. Again, for notational convenience, $p=1$ for $X_r, X_c, X_d$.  Let $X$ be a $n$-dimensional vector and define a dyadic covariate as $X_d = XX^t$. To allow for community dependence, define:
\begin{equation*}
\diag(\tilde{\boldsymbol\beta}_{dr}U^t)X_d\diag(V \tilde{\boldsymbol\beta}_{dc}^t)= \diag(\tilde{\boldsymbol\beta}_{dr}U^t)XX^t\diag( \tilde{\boldsymbol\beta}_{dc}V^t).
\end{equation*}
The full model then becomes:
\begin{equation*}\label{matrixFullModelDyad}
\begin{array}{l}
Z = \beta_0 \mathbf{11}^t + \sum_{l=1}^p (\diag(\tilde{\boldsymbol\beta}_{rl}U^t) X_{rl} + X_{cl} \diag(\tilde{\boldsymbol\beta}_{cl}V^t) +\\\diag(U \tilde{\boldsymbol\beta}_{dr,l}^t)X_d\diag(V \tilde{\boldsymbol\beta}_{dc,l}^t)) + U\Lambda V^t + \mathbf{a1}^t+ \mathbf{1b}^t + E.
\end{array}
\end{equation*}
Given the special structure of the dyadic covariate, the following holds: 
\begin{equation*}
\diag(\tilde{\boldsymbol\beta}_{dr}U^t)X =   \diag(X)U\tilde{\boldsymbol\beta}_{dr}^t.
\end{equation*}
Thus, by substitution:
\begin{align*}
\diag(\tilde{\boldsymbol\beta}_{dr}U^t)XX^t\diag(\tilde{\boldsymbol\beta}_{dc}V^t) &= \diag(X)U\tilde{\boldsymbol\beta}_{dr}^tX^t\diag( \tilde{\boldsymbol\beta}_{dc}V^t)\\
&= \diag(\tilde{\boldsymbol\beta}_{dr}U^t)X [\tilde{\boldsymbol\beta}_{dc}V^t \diag(X)].
\end{align*}
To derive full conditionals, vectorize the dyadic term to isolate $\tilde{\beta}_{dr},\tilde{\beta}_{dc}$:
\begin{equation*}
vec(\diag(X)U\tilde{\boldsymbol\beta}_{dr}^tX^t \diag(\tilde{\boldsymbol\beta}_{dc}V^t)= \{[X^t\diag(\tilde{\boldsymbol\beta}_{dc}V^t)]^t \otimes \diag(X)U\} vec(\tilde{\boldsymbol\beta}_{dr})
\end{equation*}
\begin{equation*}
vec(\diag(\tilde{\boldsymbol\beta}_{dr}U^t)X [\tilde{\boldsymbol\beta}_{dc}V^t \diag(X)])= \{ \diag(X)V \otimes \diag( \tilde{\boldsymbol\beta}_{dr}U^t)X\} vec(\tilde{\boldsymbol\beta}_{dc}).
\end{equation*}
Define:
\begin{equation*}
\tilde{Z} = Z- \beta_0 \mathbf{11}^t -  (\diag(\tilde{\boldsymbol\beta}_{r}U^t) X_{r} + X_{c} \diag(\tilde{\boldsymbol\beta}_{c}V^t))- U\Lambda V^t - \mathbf{a1}^t- \mathbf{1b}^t ,
\end{equation*}
which is decorrelated:
\begin{equation*}
\tilde{Z}_{*} = s\times \tilde{Z} + t\times \tilde{Z}^t 
\end{equation*}

where
\begin{equation*}\label{eq:HcD_dyad}
H_{dc} = s\times \{\diag(X)V \otimes \diag( \tilde{\beta}_{dr}U^t)X\} + t \times  \{[\diag(\tilde{\boldsymbol\beta}_{dr}U^t)X] \otimes \diag(X)V\},
\end{equation*}
\begin{equation*}\label{eq:HrD_dyad}
H_{dr} = s\times \{\diag(\tilde{\boldsymbol\beta}_{dc}V^t)X \otimes \diag(X)U\}+ t\times\{\diag(X)U \otimes \diag( \tilde{\boldsymbol\beta}_{dc}V^t)X\} .
\end{equation*}
The full conditionals for $\tilde{\boldsymbol\beta}_{dc}$ and $\tilde{\boldsymbol\beta}_{dr}$ are:
\begin{equation*}\label{eq:condBetaR_dyad}
\tilde{\beta}_{dr}| X_{c}, X_{r}, U, \Lambda, Z, \tilde{\beta}_{dc}, \mathbf{a},\mathbf{b}\sim N(V_{dr}\mathbf{m}_{dr}, V_{dr})
\end{equation*}
\begin{equation*}\label{eq:condBetaC_dyad}
\tilde{\beta}_{dc}| X_{c}, X_{r}, U, \Lambda, Z, \tilde{\beta}_{dr}, \mathbf{a},\mathbf{b}\sim N(V_{dc}\mathbf{m}_{dc}, V_{dc}).
\end{equation*}
where
\begin{equation*}
\begin{split}
V_{dc} &= (H_{dc}^t H_{dc} + \Sigma_{\tilde{\boldsymbol\beta}_{dc}}^{-1})^{-1}\\
m_{dc} &=vec(\tilde{Z}_{*})^tH_{dc} +\boldsymbol\mu_{\tilde{\boldsymbol\beta}_{dc}} \Sigma_{\tilde{\boldsymbol\beta}_{dc}}^{-1}
\end{split}
\quad \quad
\begin{split}
V_{dr} &= (H_{dr}^t H_{dr}+ \Sigma_{\tilde{\boldsymbol\beta}_{dr}}^{-1})^{-1}\\
m_{dr} &=vec(\tilde{Z}_{*})^tH_{dr} + \boldsymbol\mu_{\tilde{\boldsymbol\beta}_{dr}} \Sigma_{\tilde{\boldsymbol\beta}_{dr}}^{-1}
\end{split}.
\end{equation*}

\section{General Dyadic Covariates}\label{generalDyad}
Consider a generalized dyadic covariate, $X_d$ that does not take on an outer product form. Rewrite $X_d = Q\Sigma W^t$ (singular value decomposition of $X_d$ where $\Sigma$ is a diagonal matrix of singular values) and thus $X_d  = \sum_{i = 1}^n \sigma_i \mathbf{q}_i \mathbf{w}_i^t$ where $\sigma_i$ are the diagonal elements of $\Sigma$. As such, 
\begin{equation*}
\diag(\tilde{\boldsymbol\beta}_{dr}U^t)X_d \diag(\tilde{\boldsymbol\beta}_{dc}V^t) =  \diag( \tilde{\boldsymbol\beta}_{dr}U^t) \sum_{i = 1}^n \sigma_i \mathbf{q}_i \mathbf{w}_i^t\diag( \tilde{\boldsymbol\beta}_{dc}V^t).
\end{equation*}
The above can be expanded to:
\begin{align*}
 \diag( \tilde{\boldsymbol\beta}_{dr}U^t)  \sigma_1 \mathbf{q}_1 \mathbf{w}_1^t \diag( \tilde{\boldsymbol\beta}_{dc}V^t)  + ... + \diag(\tilde{\boldsymbol\beta}_{dr}U^t)  \sigma_n \mathbf{q}_n \mathbf{w}_n^t \diag(\tilde{\boldsymbol\beta}_{dc}V^t).
\end{align*}
Isolate $\tilde{\boldsymbol\beta}_{dc}$:
\begin{equation*}
\diag(\tilde{\boldsymbol\beta}_{dr}U^t)  \sigma_1 \mathbf{q}_1 \mathbf{w}_1^t \diag( \tilde{\boldsymbol\beta}_{dc}V^t) = \diag( \tilde{\boldsymbol\beta}_{dr}U^t) \sigma_1 \mathbf{q}_1 [\tilde{\boldsymbol\beta}_{dc}V^t \diag(\mathbf{w}_1)]
\end{equation*}
and vectorize to get:
\begin{equation*}
vec(\diag( \tilde{\boldsymbol\beta}_{dr}U^t)X_d [\tilde{\boldsymbol\beta}_{dc}V^t \diag(X)])= \{\diag(\mathbf{w}_1)V \otimes \diag( \tilde{\boldsymbol\beta}_{dr}U^t) \sigma_1 \mathbf{q}_1\} vec(\tilde{\boldsymbol\beta}_{dc}).
\end{equation*}
For $\tilde{\boldsymbol\beta}_{dr}$,
\begin{equation*}
\diag( \tilde{\boldsymbol\beta}_{dr}U^t)\sigma_1 \mathbf{q}_1 \mathbf{w}_1^t \diag(\tilde{\boldsymbol\beta}_{dc}V^t) = \diag(\sigma_1 \mathbf{q}_1)U\tilde{\boldsymbol\beta}_{dr}^t \mathbf{w}_1^t\diag(V \tilde{\boldsymbol\beta}_{dc}^t),
\end{equation*}
which can be vectorized as
\begin{equation*}
vec(\diag( \tilde{\boldsymbol\beta}_{dr}U^t)\sigma_1 \mathbf{q}_1 \mathbf{w}_1^t \diag(\tilde{\boldsymbol\beta}_{dc}V^t))= \{[\mathbf{w}_1^t\diag(V \tilde{\boldsymbol\beta}_{dc}^t)]^t \otimes \diag(\sigma_1\mathbf{q}_1)U\} vec(\tilde{\boldsymbol\beta}_{dr}).
\end{equation*}
Now, to derive the full conditionals, first define analogues to equations (9) and (10) from the Section~3.2.1:
\begin{equation*}\label{eq:HcD_dyadfull}
H_{dc} = s\times \sum_{i=1}^n \{\diag(\mathbf{w}_i)V \otimes \diag( \tilde{\boldsymbol\beta}_{dr}U^t) \sigma_i \mathbf{q}_i\} + t\times\sum_{i=1}^n \{[\mathbf{w}_i^t\diag( \tilde{\boldsymbol\beta}_{dr}U^t)]^t \otimes \diag(\sigma_i \mathbf{q}_i)V\},
\end{equation*}
and
\begin{equation*}\label{eq:HrD_dyadfull}
H_{dr} = s\times \sum_{i=1}^n \{[\mathbf{w}_i^t\diag(\tilde{\boldsymbol\beta}_{dc}V^t)]^t \otimes \diag(\sigma_i \mathbf{q}_i)U\}+ t \times \sum_{i=1}^n \{\diag(\mathbf{w}_i)U \otimes \diag( \tilde{\boldsymbol\beta}_{dc}V^t) \sigma_i \mathbf{q}_i\}.
\end{equation*}
 The full conditionals for $\tilde{\boldsymbol\beta}_{dr}$ and $\tilde{\boldsymbol\beta}_{dc}$ are given by:
\begin{equation*}\label{eq:condBetaR_dyadfull}
\tilde{\boldsymbol\beta}_{dr}| X_{c}, X_{r}, U, \Lambda, Z, \tilde{\boldsymbol\beta}_{dc}, \mathbf{a},\mathbf{b}\sim N(V_{dr}\mathbf{m}_{dr}, V_{dr})
\end{equation*}
\begin{equation*}\label{eq:condBetaC_dyadfull}
\tilde{\boldsymbol\beta}_{dc}| X_{c}, X_{r}, U, \Lambda, Z, \tilde{\boldsymbol\beta}_{dr}, \mathbf{a},\mathbf{b}\sim N(V_{dc}\mathbf{m}_{dc}, V_{dc})
\end{equation*}
where 
\begin{equation*}
\begin{split}
V_{dc} &= (H_{dc}^t H_{dc} + \Sigma_{\tilde{\beta}_{dc}}^{-1})^{-1}\\
\mathbf{m}_{dc} &=vec(\tilde{Z}_{*})^tH_{dc} + \boldsymbol\mu_{\tilde{\boldsymbol\beta}_{dc}} \Sigma_{\tilde{\beta}_{dc}}^{-1}
\end{split}
\quad \quad
\begin{split}
V_{dr} &= (H_{dr}^t H_{dr}+ \Sigma_{\tilde{\beta}_{dr}}^{-1})^{-1}\\
m_{dr} &=vec(\tilde{Z}_{*})^tH_{dr} + \boldsymbol\mu_{\tilde{\boldsymbol\beta}_{dr}} \Sigma_{\tilde{\beta}_{dr}}^{-1}
\end{split}
\end{equation*}

\section{Bivariate Probit Likelihood}\label{probitLik}
 Let $m_{ij} =  \beta_0+ \sum_{l=1}^p (x_{irl} \tilde{\boldsymbol\beta}_{rl} \mathbf{u}_i^t + x_{jcl}\tilde{\boldsymbol\beta}_{cl}\mathbf{v}_j^t) + a_i + b_j  + \mathbf{u}_i^t \Lambda \mathbf{v}_j$. The likelihood is then: 
\begin{equation}
\begin{array}{l}
\log(L(\beta, U,V, \Lambda, \mathbf{a},\mathbf{b}|X,Y))=\sum_{i <j}(y_{ij}y_{ji}\log(\Phi(m_{ij} , m_{ji} , \rho))+\\
(1-y_{ij})y_{ji}\ log(\Phi(-(m_{ij} ), m_{ji} , -\rho)) +\\ 
y_{ij}(1-y_{ji}\log(\Phi(m_{ij}, -(m_{ji}), -\rho)) +\\
(1-y_{ij})(1-y_{ji}) \log(\Phi(-(m_{ij}), -(m_{ji}), \rho)) 
) +\\\sum_{i=1}^n \log(1-\Phi(m_{ii} , \sqrt{1+\rho}))
\end{array}
\end{equation}
where $\Phi(x,y, \rho)$ indicates evaluating the cumulative distribution function of the bivariate normal distribution.

\section{Comparison of Censored Binary formulations} \label{compare_censor}

\begin{figure}[H]
	\centering
	\includegraphics[width=.8\columnwidth]{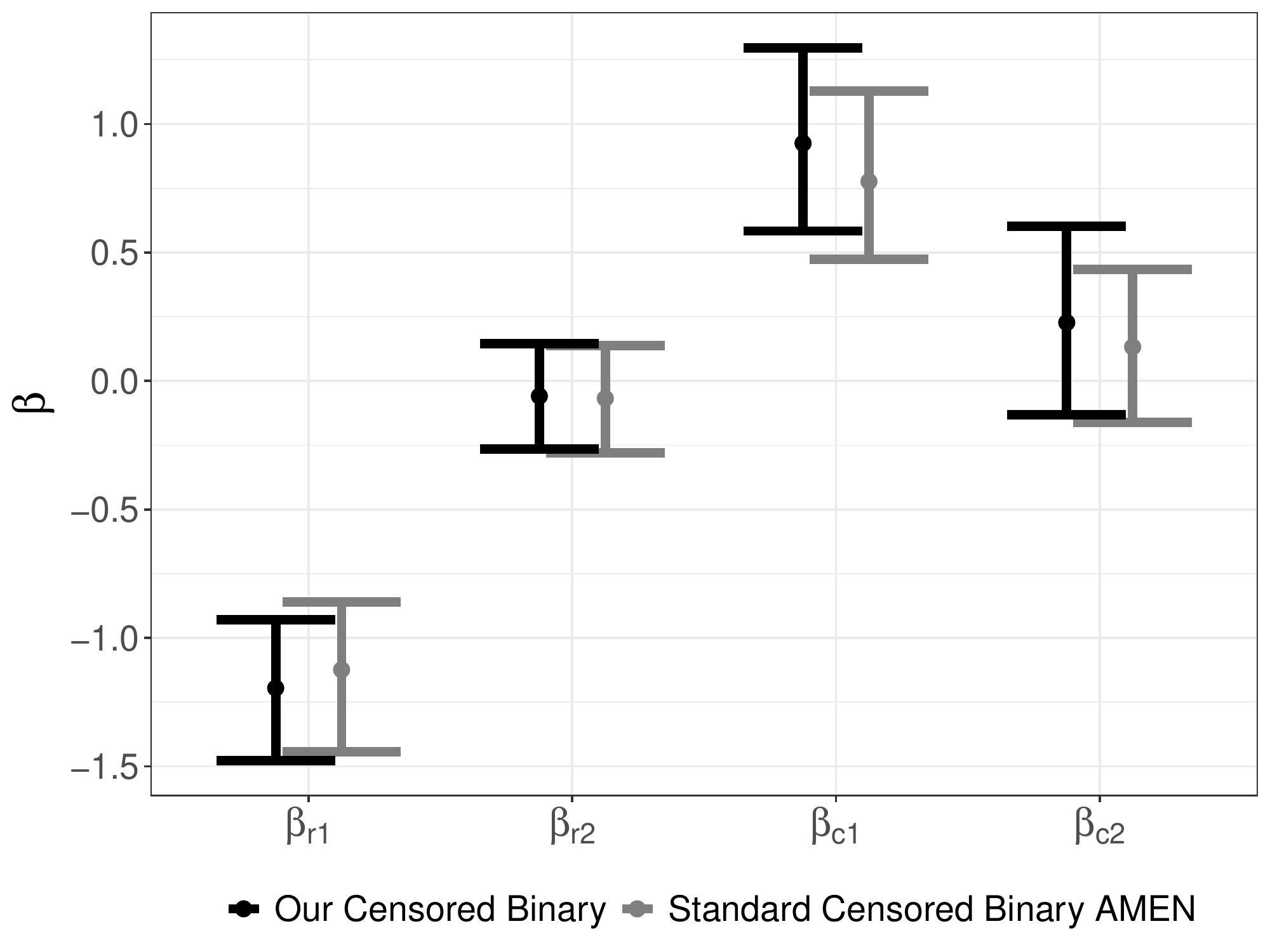}
	\caption{$95\%$ CI for row and column coefficients under our version of the censored binary versus that of the standard censored binary AMEN model as in the 'amen' package. Data is generated as in Section 4.3.  \label{fig:comparing_censored}}
\end{figure}

A censored binary likelihood for networks was introduced in Hoff (2013), leveraging the set-based likelihood machinery of that paper. The formulation in that work does not naturally lend itself to our proposed updates for community dependent coefficients. In this section we demonstrate that the two approaches lead to the same parameter estimates (in a non-community dependent setting). The standard version of the model matches the data generating mechanism for the censored network structure as in Section 4.3.  That is, once $Z$ values are simulated, $y_{ij} = 1$ if $z_{ij}$ is in the top $m$ $Z_{i.}$ values. Both models yield near equivalent posterior estimates for all parameters of interest as shown in Figure \ref{fig:comparing_censored}. As an additional note, recall that in Section 4.3, there are community dependent covariate effects that are not accounted for in these non-community dependent models. Specifically, $\beta_{r2}$ and $\beta_{c2}$ vary between communities. The posterior estimates for these coefficients are approximately $0$ for both estimation approaches (which is the average of the community dependent $\tilde{\boldsymbol\beta}$ values). 

\section{Covariate Information for AddHealth}\label{covar_addhealth}
In the AddHealth study, high school students (grades 9-12) ranked their top $5$ female and top $5$ male friends. We analyze the same-sex male-to-male friendships in two schools. Covariates of interest include:
\begin{itemize}
    \item Race --- an indicator of whether the individual is white
    \item Drinking --- an indicator of whether a an individual drinks
    \item Smoking --- an indicator of whether an individual smokes
    \item GPA --- numeric grade point average
    \item Living situation
    \begin{itemize}
        \item Living with father
        \item Living with mother
        \item Number of people in household
    \end{itemize}
    \item Club involvement --- number of clubs a student is involved in 
    \item Sport involvement --- number of sports a student is involved in 
\end{itemize}
 
We further include dyadic covariates for whether or not two students share the same grade, share the same binary drinking behavior, share the same binary smoking behavior, and whether they share the same race.  

\section{Trace plots for AddHealth Networks}\label{traceplot}
\begin{figure}[!h]
    \centering
    \includegraphics[width = .8\columnwidth ]{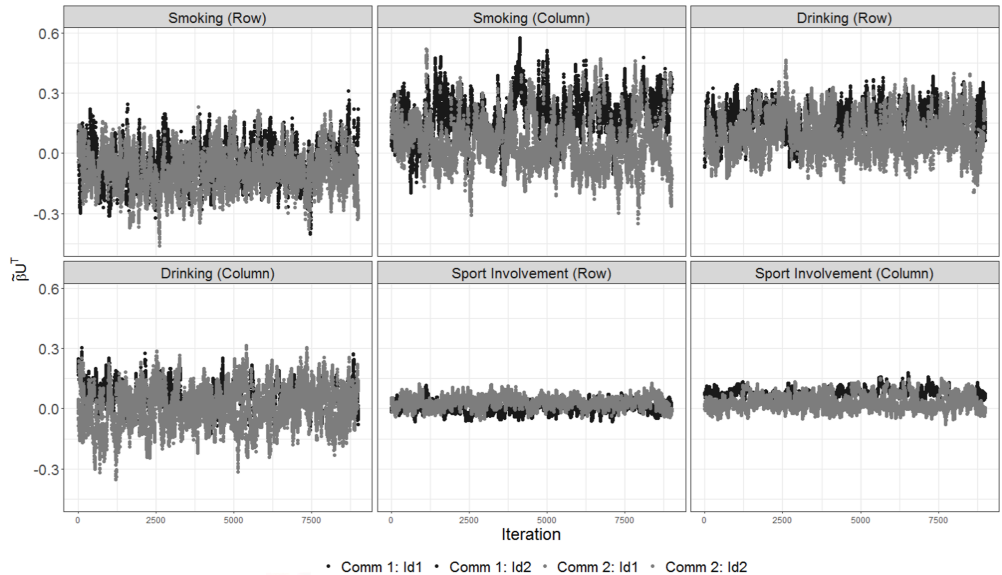}
    \caption{Trace plot for $4$ individuals' $u_i \tilde{\boldsymbol\beta}$ for multiple covariates from school B. The nodes belonging to the same community overlap so two distinct colors are used to distinguish between communities.}
    \label{fig:traceplot}
\end{figure}

Geweke $z$-statistics assessing the convergence of the Markov Chain Monte Carlo algorithm are reported in the main text of the article. Figure~\ref{fig:traceplot} presents traceplots for several parameters in the model for visual inspection of the stationarity of the chain. 
\section{Dyadic Posterior Credible Intervals for AddHealth Networks}\label{addhealth_dyad}

\begin{figure}[H]
	\centering
	\includegraphics[width=.8\columnwidth]{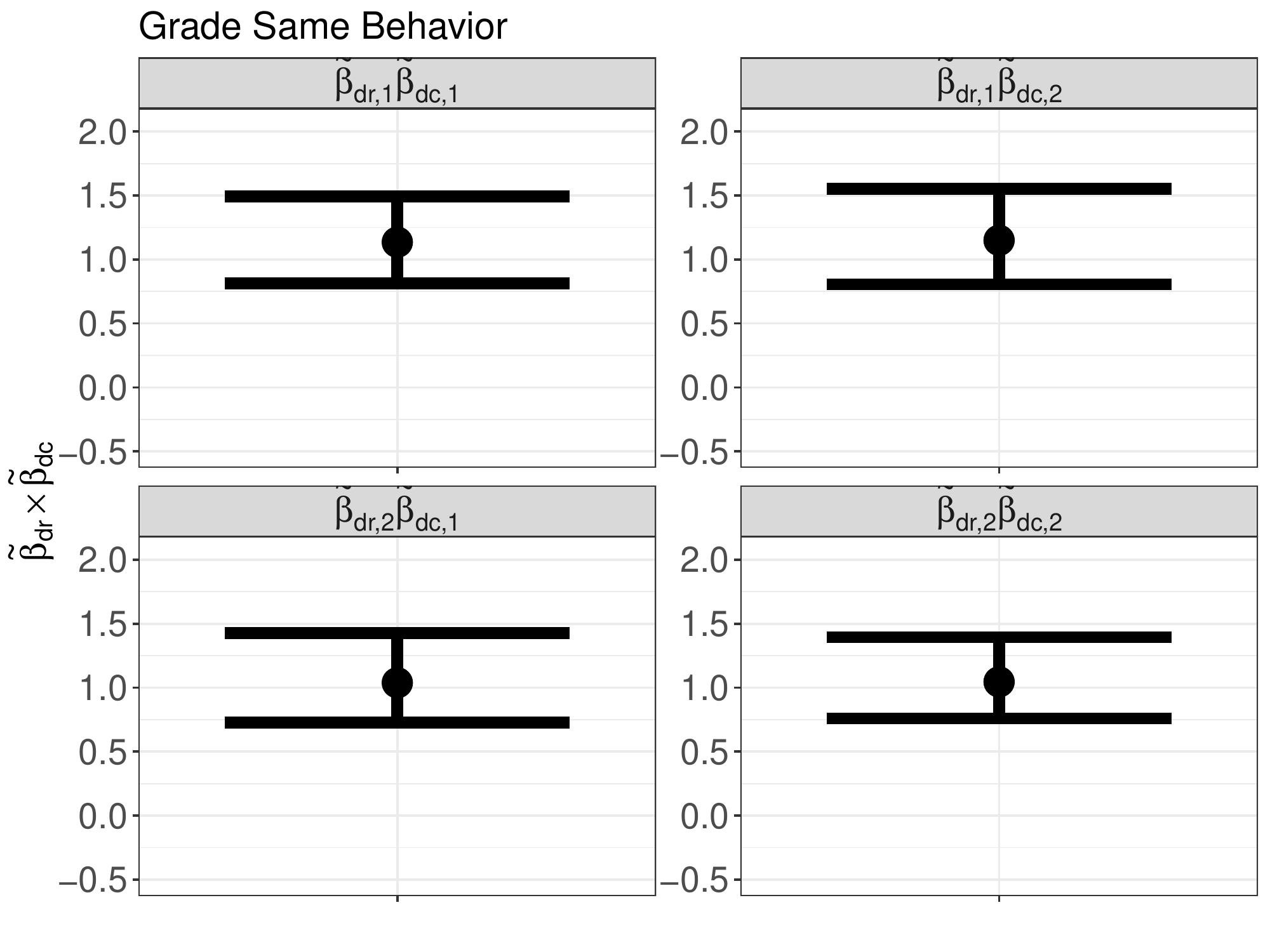}
	\caption{$95\%$ CI for dyadic covariates with community dependent model, $K = 2$ \label{fig:addHealth_grade22}}
\end{figure}

\begin{figure}[H]
	\centering
	\includegraphics[width=.8\columnwidth]{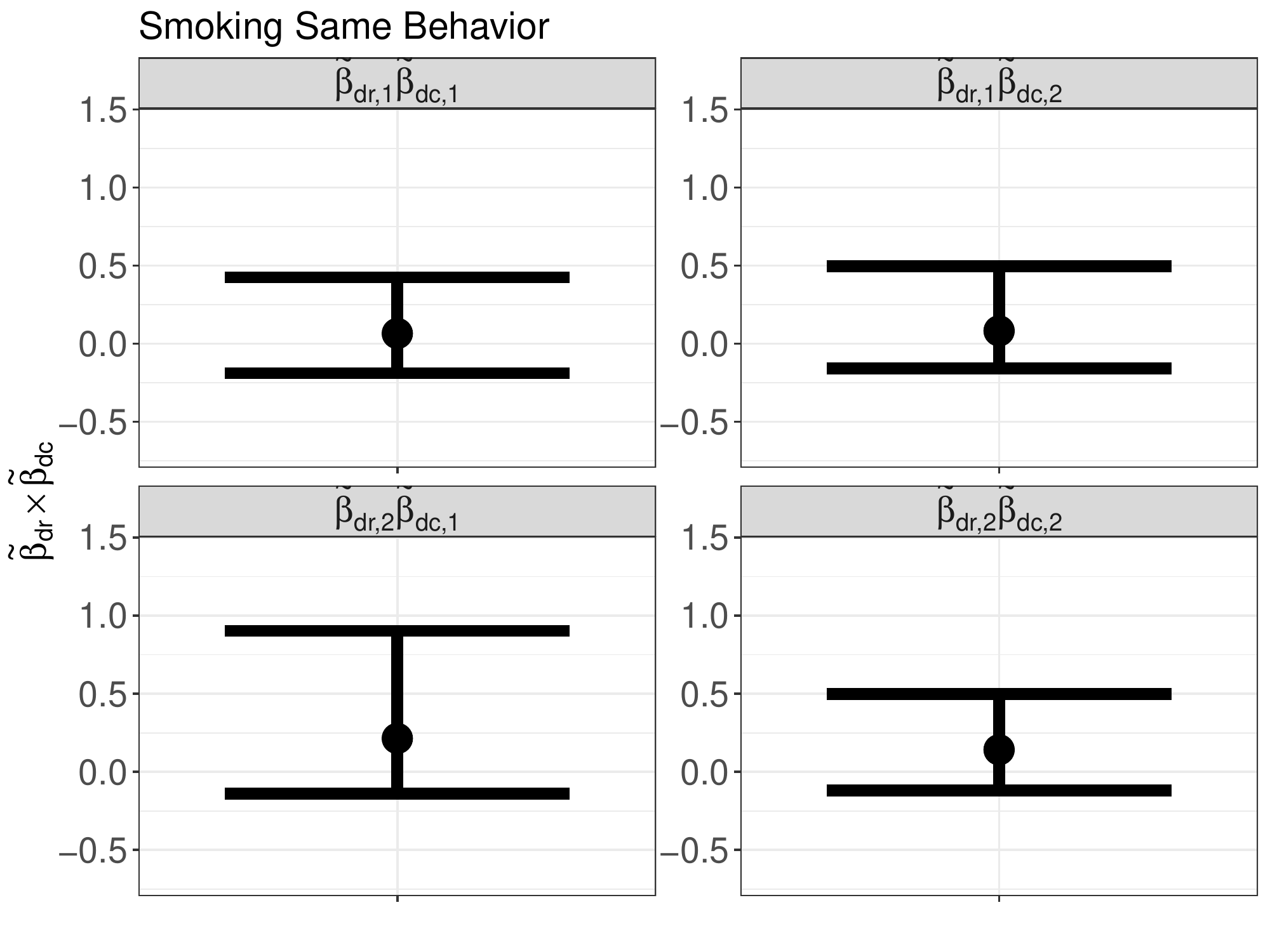}
	\caption{$95\%$ CI for dyadic covariates with community dependent model, $K = 2$ \label{fig:addHealth_smoking22}}
\end{figure}

\begin{figure}[H]
	\centering
	\includegraphics[width=.8\columnwidth]{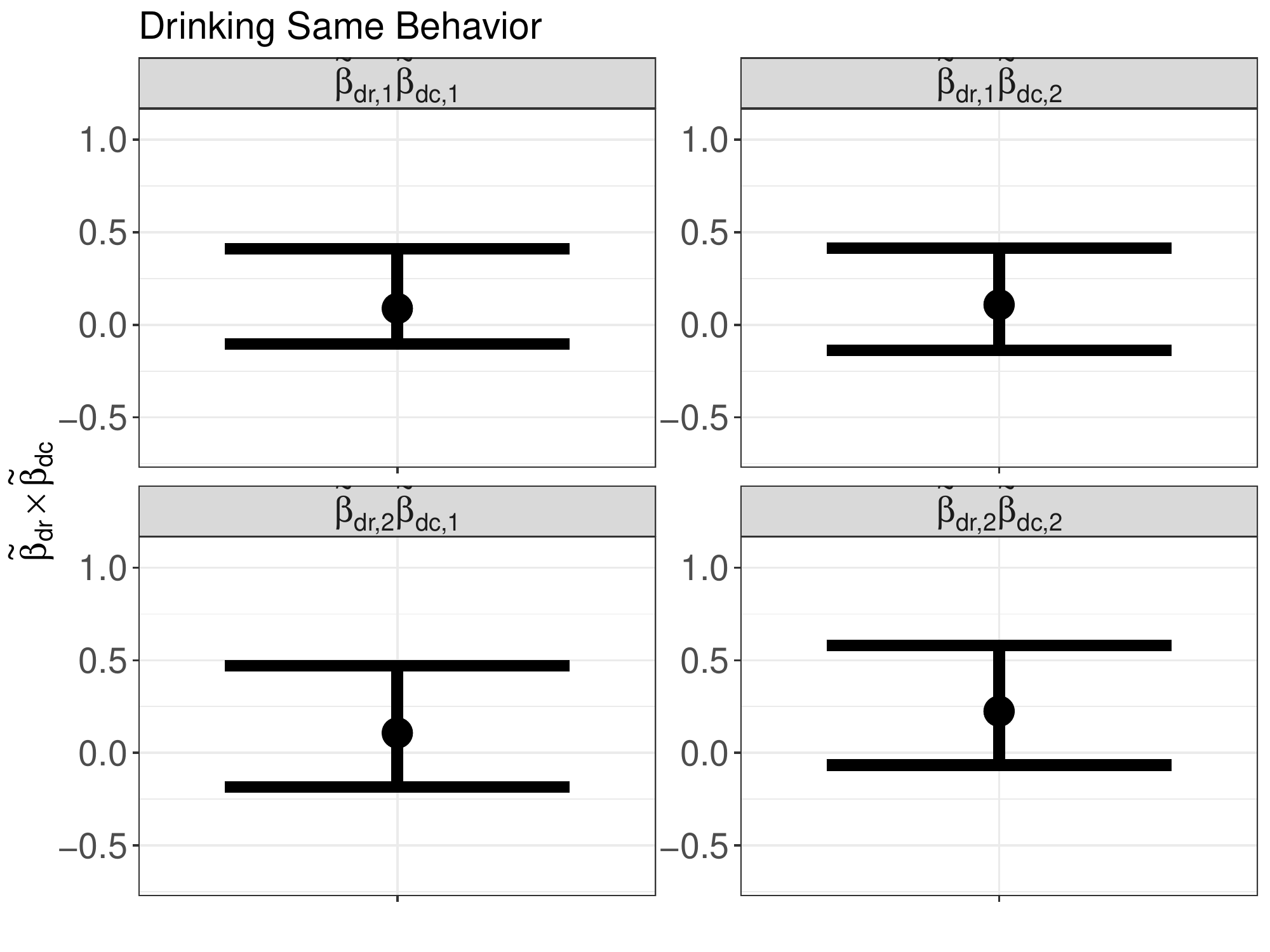}
	\caption{$95\%$ CI for dyadic covariates with community dependent model, $K = 2$ \label{fig:addHealth_drinking22}}
\end{figure}

\begin{figure}[H]
	\centering
	\includegraphics[width=.8\columnwidth]{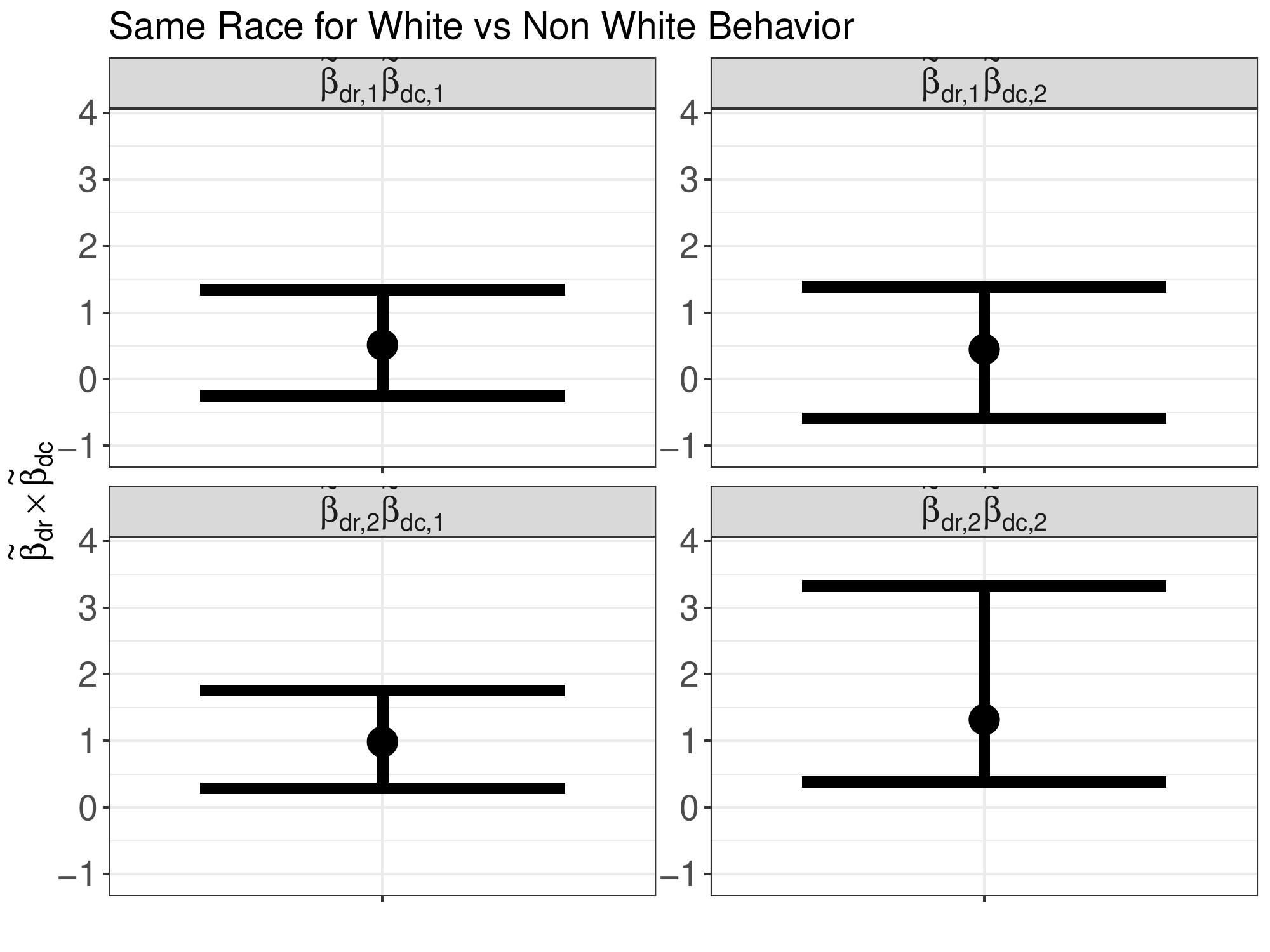}
	\caption{$95\%$ CI for dyadic covariates with community dependent model, $K = 2$ \label{fig:addHealth_race22}}
\end{figure}

\begin{figure}[H]
	\centering
	\includegraphics[width=.8\columnwidth]{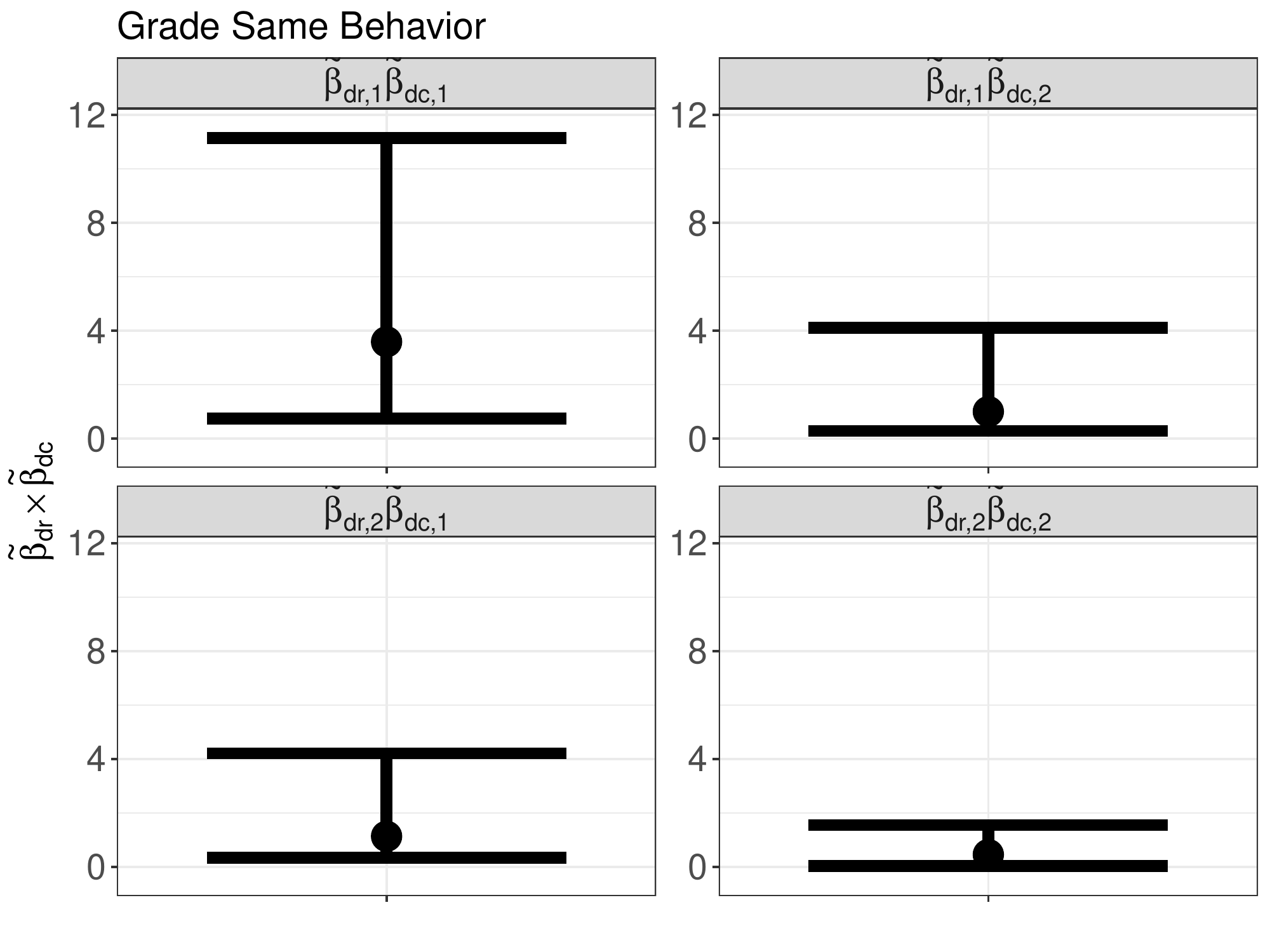}
	\caption{$95\%$ CI for dyadic covariates with community dependent model, $K = 2$ \label{fig:addHealth_grade23}}
\end{figure}

\begin{figure}[H]
	\centering
	\includegraphics[width=.8\columnwidth]{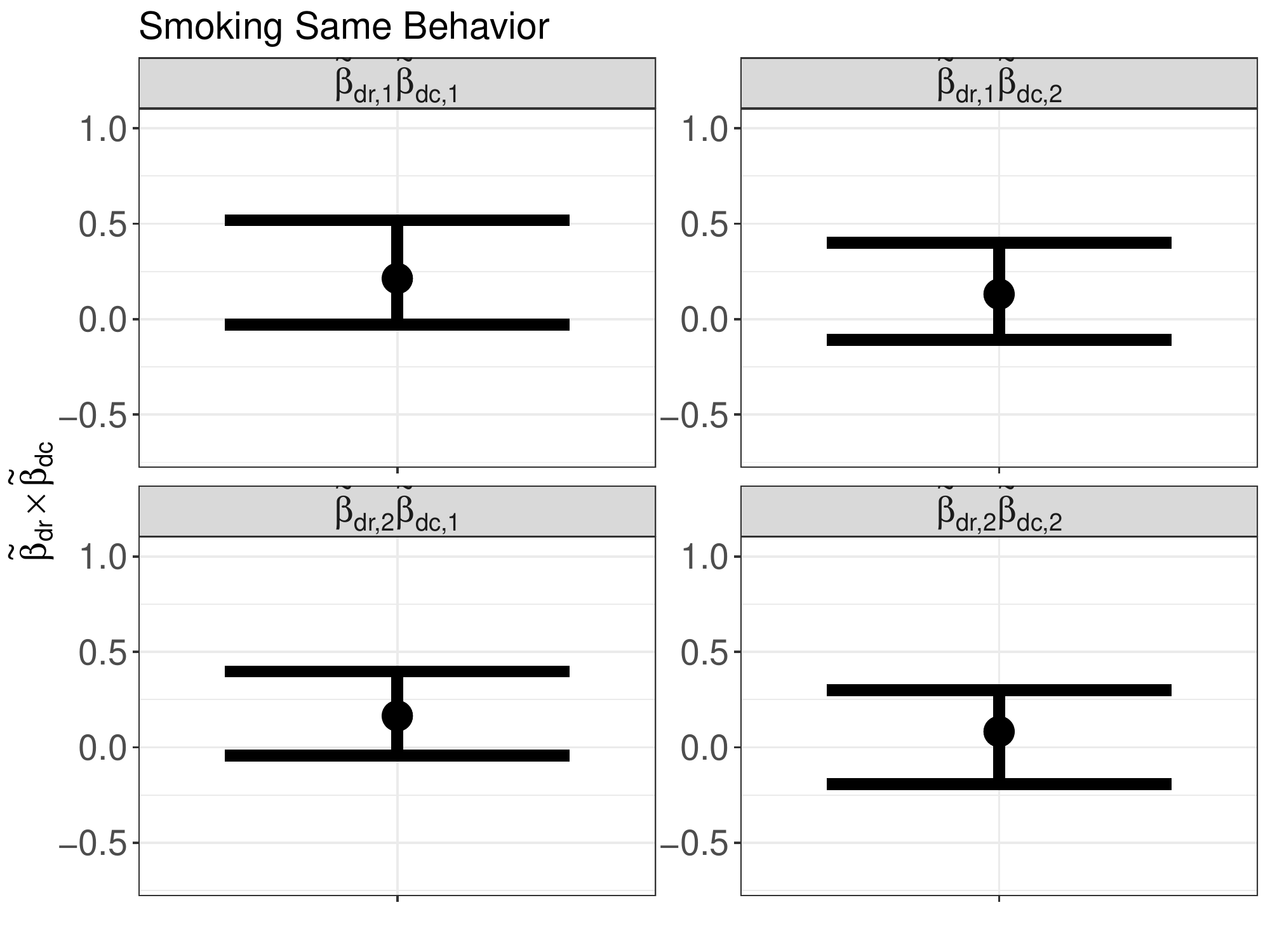}
	\caption{$95\%$ CI for dyadic covariates with community dependent model, $K = 2$ \label{fig:addHealth_smoking23}}
\end{figure}

\begin{figure}[H]
	\centering
	\includegraphics[width=.8\columnwidth]{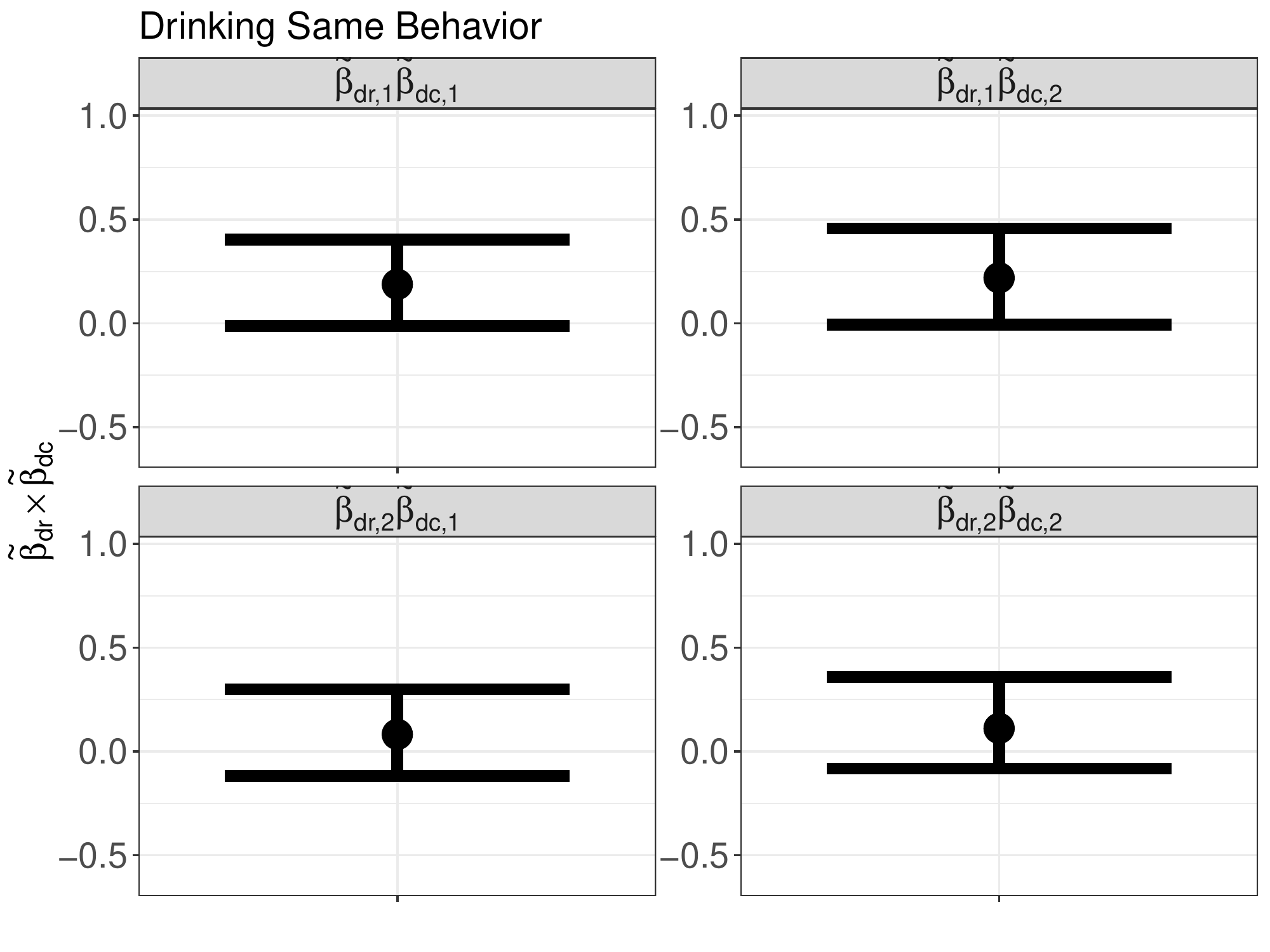}
	\caption{$95\%$ CI for dyadic covariates with community dependent model, $K = 2$ \label{fig:addHealth_drinking23}}
\end{figure}

\begin{figure}[H]
	\centering
	\includegraphics[width=.8\columnwidth]{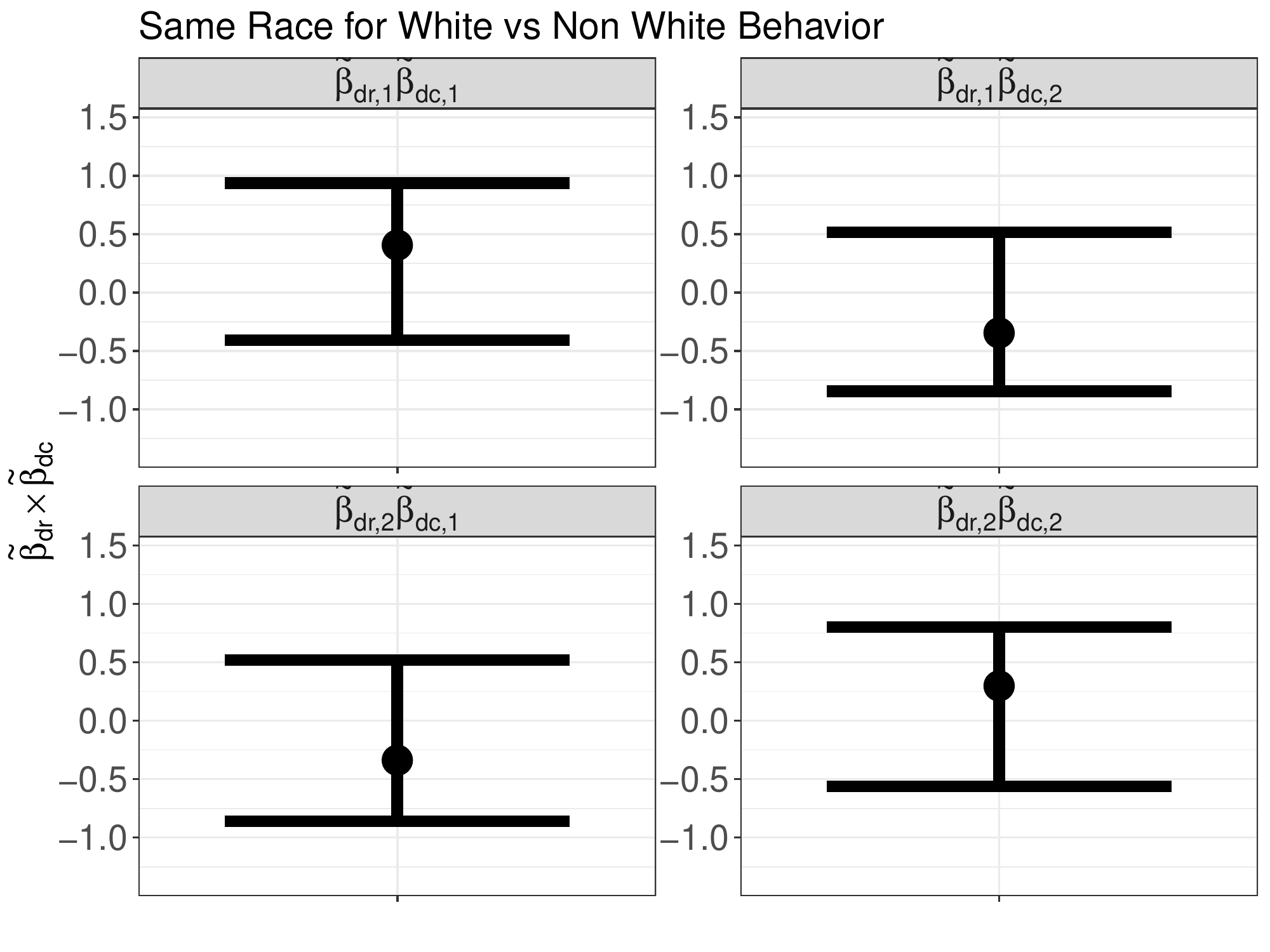}
	\caption{$95\%$ CI for dyadic covariates with community dependent model, $K = 2$ \label{fig:addHealth_race23}}
\end{figure}

Each of Figures~\ref{fig:addHealth_grade22}-\ref{fig:addHealth_race23} present the post-processed posterior credible intervals for the community dependent dyadic covariates. In each figure, the top left panel (labeled $\tilde{\beta}_{dr,1}\tilde{\beta}_{dc,1}$) represents the coefficient for a pair of individuals where both sender and receiver are in community 1. The top right panel (labeled $\tilde{\beta}_{dr,1}\tilde{\beta}_{dc,2}$) represents the coefficient for a pair of individuals where the sender is in community 1 and the receiver is in community 2. The bottom left figure (labeled $\tilde{\beta}_{dr,2}\tilde{\beta}_{dc,1}$) represents the coefficient for a pair of individuals where the sender is in community 2 and the receiver is in community 1. The bottom right panel (labeled $\tilde{\beta}_{dr,2}\tilde{\beta}_{dc,2}$) represents the coefficient for a pair of individuals where both sender and receiver are in community 2.

\subsection{School A}
The following figures show the $95\%$ credible intervals for school A for dyadic covariates. As with the row and column effects that we present in the main paper, the dyadic covariates for this school exhibit little to no community structure.

\subsection{School B}
The following figures show the $95\%$ credible intervals for school B for dyadic covariates. We note that community dependent behavior is exhibited for the smoking and drinking behavior: pairs where both sender and receiver are in community 1 have positive effects on friendship formation when both are smokers or drinkers. This behavior is completely absent for smoking when both sender and receiver are in community 2 and is largely absent for drinking when at least the sender of ties is in community 2.  

\end{document}